\documentclass[useAMS,usenatbib]{mn2e}

\usepackage{array} %Líneas gruesas en tablas
\usepackage{multirow} %Permite unir filas en las tablas.
\usepackage{booktabs} %Permite embellecer tablas.
\usepackage{tabularx} %Permite embellecer tablas.

\usepackage{gensymb}% Degree symbol
\usepackage{graphicx}
\usepackage[english]{babel}
\usepackage{txfonts}
\usepackage{multirow}
\usepackage{lscape}
\usepackage{longtable}
\usepackage{natbib}
\usepackage{bibentry} \usepackage{color}
\usepackage{float}
\usepackage{array}
\usepackage{graphicx,amsfonts,psfrag,fancyhdr,layout,appendix}%,subfigure}
\usepackage{makeidx}
\usepackage{fancyvrb}
\usepackage[labelformat=simple]{caption,subcaption}
\usepackage{multirow}
\usepackage{dcolumn}
\usepackage{hyperref,xcolor}%[hidelinks]
\hypersetup{
  colorlinks,
  citecolor=blue,
  linkcolor=blue,
  urlcolor=blue}
\usepackage{xcolor}
\usepackage{etoolbox}

\makeatletter

\patchcmd{\NAT@citex}
  {\@citea\NAT@hyper@{%
     \NAT@nmfmt{\NAT@nm}%
     \hyper@natlinkbreak{\NAT@aysep\NAT@spacechar}{\@citeb\@extra@b@citeb}%
     \NAT@date}}
  {\@citea\NAT@nmfmt{\NAT@nm}%
   \NAT@aysep\NAT@spacechar\NAT@hyper@{\NAT@date}}{}{}

\patchcmd{\NAT@citex}
  {\@citea\NAT@hyper@{%
     \NAT@nmfmt{\NAT@nm}%
     \hyper@natlinkbreak{\NAT@spacechar\NAT@@open\if*#1*\else#1\NAT@spacechar\fi}%
       {\@citeb\@extra@b@citeb}%
     \NAT@date}}
  {\@citea\NAT@nmfmt{\NAT@nm}%
   \NAT@spacechar\NAT@@open\if*#1*\else#1\NAT@spacechar\fi\NAT@hyper@{\NAT@date}}
  {}{}

\makeatother

\newcommand{\lines}{$\lambda\lambda\ 4960, 5008\, \AA$}
\newcommand{\linesNe}{$\lambda\lambda\ 3869, 3968\, \AA$}
\newcommand{\linesSi}{$\lambda\lambda\ 6719, 6733\, \AA$}
\newcommand{\oxygen}{\mbox{[O {\sc iii}]}\ }
\newcommand{\neon}{\mbox{[Ne {\sc iii}]}\ }

\title[Fine-structure constant with BOSS]{Constraint on the time variation of the fine-structure constant\\ with the SDSS-III/BOSS DR12 quasar sample}

\author[F. D. Albareti et al.]{\parbox[t]{\textwidth}{Franco D. Albareti,$^{1}$\thanks{`la Caixa'-Severo Ochoa Scholar.}\thanks{E-mail: franco.albareti$@$uam.es} Johan Comparat,$^{1}$ Carlos M. Guti\'errez,$^{2,3}$ Francisco Prada,$^{1,4,5}$\\Isabelle P\^aris,$^{6}$ David Schlegel,$^{7}$ Mart\'in L\'opez-Corredoira,$^{2,3}$ Donald P. Schneider,$^{8,9}$\\ Arturo Manchado,$^{2,3,10}$ D.\,A.\ Garc\'ia-Hern\'andez,$^{2,3}$ Patrick Petitjean$^{11}$ and Jian Ge$^{12}$}\vspace*{10pt} \\
$^1$ Instituto de F\'{\i}sica Te\'orica (UAM/CSIC), Universidad Aut\'onoma de Madrid,  Cantoblanco, E-28049 Madrid, Spain \\
$^{2}$ Instituto de Astrof{\'\i}sica de Canarias (IAC), La Laguna, E-38205 Tenerife, Spain\\
$^{3}$ Departamento de Astrof\'isica, Universidad de La Laguna, La Laguna, E-38206 Tenerife, Spain\\
$^{4}$ Campus of International Excellence UAM+CSIC, Cantoblanco, E-28049 Madrid, Spain\\
$^{5}$ Instituto de Astrof\'{\i}sica de Andaluc\'{\i}a (CSIC), Glorieta de la Astronom\'{\i}a, E-18080 Granada, Spain\\
$^{6}$ INAF, Osservatorio Astronomico di Trieste, Via G.
B. Tiepolo 11, 34131 Trieste, Italy\\
$^{7}$ Lawrence Berkeley National Laboratory, 1 Cyclotron Road, Berkeley, CA, 94720, USA\\
$^{8}$ Department of Astronomy and Astrophysics, The Pennsylvania State University, University Park, PA 16802, USA\\
$^{9}$ Institute for Gravitation and the Cosmos, The Pennsylvania State University, University Park, PA 16802, USA\\
$^{10}$ CSIC, Spain\\
$^{11}$ Institut d'Astrophysique de Paris, CNRS-UPMC, UMR7095, 98bis bd Arago, 75014 Paris, France\\
$^{12}$ Department of Astronomy, University of Florida, Gainesville, FL 32611-2055, USA}

\begin{document}

\date{Accepted for publication in MNRAS.}

\pagerange{\pageref{firstpage}--\pageref{lastpage}} \pubyear{2002}

\maketitle

\label{firstpage}

%%%%%%%%%%%%%%%%%%%%%%%%%%%%%%%%%%%%%%%%%%%%%%%%%%%%%%
%%%%%%%%%%%%%%%%%%%%%%%%%%%%%%%%%%%%%%%%%%%%%%%%%%%%%%

\begin{abstract}

From the Sloan Digital Sky Survey (SDSS) Data Release 12, which covers the full Baryonic Oscillation Spectroscopic Survey (BOSS) footprint, we investigate the possible variation of the fine-structure constant over cosmological time-scales. 
We analyse the largest quasar sample considered so far in the literature, which contains 13\,175 spectra (\mbox{10\,363 from SDSS-III/BOSS} DR12 + 2812 from SDSS-II DR7) with redshift $z<1$. We apply the emission-line method on the \oxygen doublet (\lines) and obtain $\Delta\alpha/\alpha= \left(0.9 \pm 1.8\right)\times10^{-5}$ for the relative variation of the fine-structure constant.
We also investigate the possible sources of systematics: misidentification of the lines, sky OH lines, H$\,\beta$ and broad line contamination, Gaussian and Voigt fitting profiles, optimal wavelength range for the Gaussian fits, chosen polynomial order for the continuum spectrum, signal-to-noise ratio and good quality of the fits. The uncertainty of the measurement is dominated by the sky subtraction. % induced systematic error.
 The results presented in this work, being systematics limited, have sufficient statistics to constrain robustly the variation of the fine-structure constant in redshift bins ($\Delta z\approx 0.06$) over the last 7.9 Gyr. In addition, we study the \neon doublet (\linesNe) present in $462$ quasar spectra and discuss the systematic effects on using these emission lines to constrain the fine-structure constant variation. Better constraints on $\Delta\alpha/\alpha\  (<10^{-6})$ using the emission-line method would be possible with high-resolution spectroscopy and large galaxy/qso surveys.

%Using a high-resolution Planetary Nebula spectrum, we further investigate the \neon (\linesNe) fine-structure doublet as a potential future-, higher-redshift alpha probe and clearly exhibit that the contamination by the H$\epsilon$ line induces a biased centroiding of the second Ne line.

%We show that the final error is dominated by the sky subtract%ion which limits the current precision. 
%??? We classify the spectra according to their quality parameters, and study different samples. ??? The \neon doublet (\linesNe)  is also analysed and we found a positive systematic effect on the value for $\Delta\alpha/\alpha$ already reported in the literature. A possible error on the wavelength separation of the \neon lines is excluded.
%Simulations of thousands of virtual spectra are carried out to test the precision of our code and the systematics of our measurements. 
\end{abstract}

%%%%%%%%%%%%%%%%%%%%%%%%%%%%%%%%%%%%%%%%%%%%%%%%%%%%%%
%%%%%%%%%%%%%%%%%%%%%%%%%%%%%%%%%%%%%%%%%%%%%%%%%%%%%%

\begin{keywords}
line: profiles -- quasars: emission lines -- cosmology: observations -- surveys -- large-scale structure of Universe.
\end{keywords}

%%%%%%%%%%%%%%%%%%%%%%%%%%%%%%%%%%%%%%%%%%%%%%%%%%%%%%
%%%%%%%%%%%%%%%%%%%%%%%%%%%%%%%%%%%%%%%%%%%%%%%%%%%%%%

\section{Introduction}

%%%%%%%%%%%%%%%%%%%%%%%%%%%%%%%%%%%%%%%%%%%%%%%%%%%%%%
%%%%%%%%%%%%%%%%%%%%%%%%%%%%%%%%%%%%%%%%%%%%%%%%%%%%%%
\newcolumntype{L}[1]{>{\raggedright\let\newline\\\arraybackslash\hspace{0pt}}m{#1}}
\newcolumntype{C}[1]{>{\centering\let\newline\\\arraybackslash\hspace{0pt}}m{#1}}
\newcolumntype{R}[1]{>{\raggedleft\let\newline\\\arraybackslash\hspace{0pt}}m{#1}}
\newcommand{\cf}{{\it cf.}}
\begin{table*}

\caption{Summary of the results obtained by recent works based on the \oxygen emission line method for the possible variations of the fine-structure constant.}
\label{tab:comparison}
\centering
\begin{tabular}{@{}L{4.25cm}R{1.7cm}clccC{2.15cm}D{,}{\pm}{1.1}@{}}%{c | c | c } 
\midrule[1.8pt] 
Reference & Quasar spectra & \multicolumn{2}{l}{SDSS release} & $z_{\rm{min}}$ & $z_{\rm{max}}$ & Time ago (Gyr)$^{(a)}$ &  \multicolumn{1}{c}{\hspace{-0.25cm}$\Delta\alpha/\alpha\ (\times 10^{-5})$} \\ \midrule[1.8pt] 

\citet{Bahcall} &  $42$ & EDR &\hspace{-0.4cm}\citep{EDR} & 0.16 & 0.80 & 7.0
 	& \hspace{-1.5cm} 7\ ,\ 14  \\ %\midrule[0.4pt]
 
 \citet{Gutierrez}  &  $1568$  & DR6 &\hspace{-0.4cm}\citep{DR6} & 0.00 & 0.80 & 7.0
 	&  2.4\ ,\ 2.5   \\ %\midrule[0.4pt]
 	
\citet{Rahmani}  &  $2347$  & DR7 &\hspace{-0.4cm}\citep{DR7} & 0.02 & 0.74 & 6.7	
 	&  -2.1\ ,\ 1.6\\ %\midrule[0.4pt]
    
This work (2015)  &  $13\,175$ & DR12 &\hspace{-0.35cm}\citep{DR12} & 0.04 & 1.00 & 7.9
 	&  0.9\ ,\ 1.8\ ^{(b)}\\ 
 	%& & (BOSS) & & & \\
 	 \midrule[1.8pt] 
 	 % Info de la muestra: SN>4*3, mu y nu < 25, otilier<1*10^-2
%
\end{tabular}\smallskip\\
\subcaptionbox*{$^{(a)}$ For a $\Lambda$CDM cosmology with $H_{0}=67.8\ \rm{km\,s}^{-1}\,\rm{Mpc}^{-1}$, $\Omega_{\rm{m}}=0.31$ and $\Omega_{\Lambda}=0.69$ from {\it Planck+WMAP}-9+BAO \citep{Planck}.

$^{(b)}$ Note: Since we have a larger sample than \citet{Gutierrez}, we expect a factor $\approx 2.5$ of improvement in the error just from purely statistical reasons. In Figs\ \ref{fig:sky} and \ref{fig:alpha}, it is shown that the error is dominated by the sky subtraction algorithm, which suggests that the performed analysis have reached the maximum precision with the available data.
%This makes our sample the largest sample used today, being almost independent from the ones of previous works.
}[\linewidth]

\end{table*}

Since Dirac's philosophical argument \citep{Dirac} against the fixed value of fundamental constants of Nature, several experiments have been performed to constrain possible variation on dimensionless constants of physical theories. Fundamental constants of physics could be thought of as parameters which enter in our description of Nature but they cannot be predicted with our current theories and should be measured. Dirac's idea is based on the unlikely fact that the most fundamental constants of the Universe have a certain fixed value (at a given energy) with no apparent relation with the real world. It is more likely that their present values are the result of a dynamical process, which had yielded the fundamental constants as they are measured today. Therefore, they should be considered as characterizing the state of the Universe \citep{Uzan2003}. There are many current theoretical frameworks which allow for such variation of the fundamental constants, for instance, string theory \citep{Maeda}, modified gravity and theories with extradimensions \citep[e.g.][]{Clifton}. Moreover, the experimental bounds on their variation have become a stringent test for those theoretical models \citep[e.g.][]{Thompson,Leal}. The most studied fundamental constants are the fine-structure constant $\alpha$, the Newton gravitational constant $G$ and the electron-to-proton mass ratio $\mu$ \citep{Uzan2003, Uzan2011,Garcia-Berro}.

The fine-structure constant governs the electromagnetic coupling between photons and charged particles $\alpha=e^2/(\hbar c)$. The current constraint on its relative variation $\Delta\alpha/\alpha$, over geological time-scales, is  $|\Delta\alpha/\alpha|<7\times10^{-8}$ up to $z\approx0.15$ (2\,Gyr ago); obtained from the Oklo phenomenon \citep[e.g.][]{Petrov}. It has also been reported $|\Delta\alpha/\alpha|<3\times10^{-7}$ up to $z\approx0.45$ ($4-5$\,Gyr ago) from meteorites \citep{Olive}; which also excludes possible variations on the scales of the Solar system. On the other hand, there are also constraints, $|\Delta\alpha/\alpha|\lesssim10^{-2}$, based on the cosmic microwave background \citep[CMB;][]{Landau, Planck} at $z\approx 1100$ and from big bang nucleosynthesis, the latter being model-dependent. By measuring fine-structure multiplets at different redshift in the absorption or emission spectra of galaxies and quasars, located at different directions in the sky, one can measure an estimate of the variation of $\alpha$ with time or space over cosmological scales.

The first measurements on the variation of $\alpha$ from astronomical observations reached an accuracy of $\Delta \alpha / \alpha \approx10^{-2}-10^{-3}$ \citep*{Savedoff, BahcallSalpeter, BahcallSchmidt, BahcallSargent}. Since then, the methodology and understanding of systematics has dramatically improved. Current measurements of absorption multiplets along the line of sight of three quasars around redshift 1.5, observed with spectral resolving power \mbox{$R\approx60\,000$} at UVES/ESO-VLT, reached the $\approx\, 5\times 10^{-6}$ level \citep{Evans}. Using emission lines, an accuracy of $\approx\, 2\times 10^{-5}$ was achieved analysing $1500-2300$ quasar spectra at $z\approx 0.6$ \citep{Gutierrez, Rahmani}, taken with the Sloan Digital Sky Survey (SDSS) $R\approx2000$ spectrograph.

%There exist several methods to measure the possible variations of the fine-structure constant through astronomical observations. However, they can be divided in two general classes depending on whether emission or absorption lines are used. 
The measurements on absorption features on a quasar spectrum are currently limited by the precision in the absolute wavelength calibration of the spectra, i.e., $50-200$ m\,s$^{-1}$ using spectra with $R\approx60,000$ \citep{Molaro, Evans, ESO}. Furthermore, the so-called many-multiplet (MM) method used in \citet{Evans}, although more precise, remains controversial as several assumptions are made, the most important one being ionization and chemical homogeneity. These assumptions may induce systematic biases on the value of $\alpha$. 

In this article, we use the method based on the \oxygen emission lines, first proposed by \citet{BahcallSalpeter}, which is less affected by systematics. In particular, there is no need for assuming ionization and chemical homogeneity, since the studied lines have the same profile (the transitions originate at the same upper energy level). Furthermore, the emission-line method suffers of much less spectral distortion, since the measurements of $\Delta\alpha/\alpha$ are done on a spectral window $\sim100$\,\AA\, as compared to $\sim1000$\,\AA\ when the MM method is used. With a large ensemble of quasars and/or using high-resolution spectroscopy, the uncertainty can be reduced significantly, and will compete with the absorption method when using high-resolution spectroscopy.

\begin{figure*}
\centerline{\includegraphics[angle=0,width=0.495\textwidth]{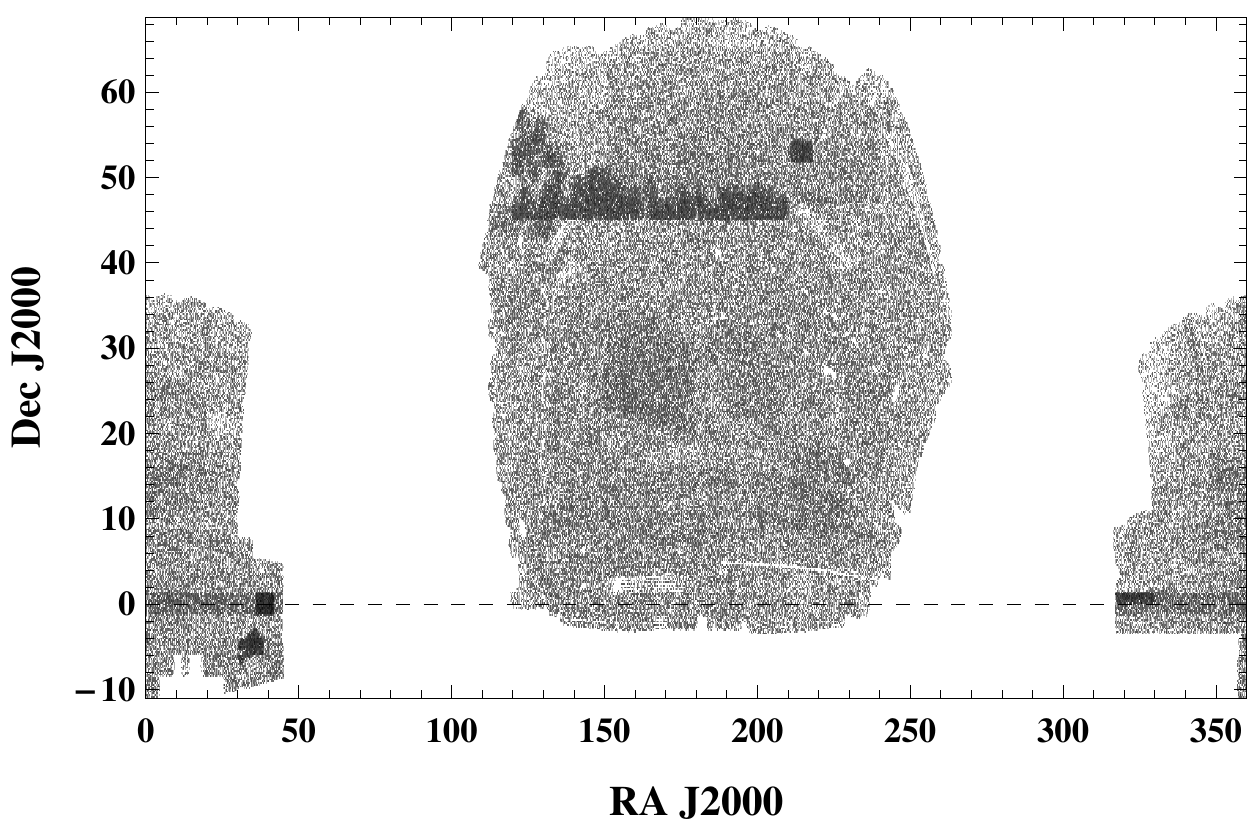}\hfill
			\includegraphics[angle=0,width=0.495\textwidth]{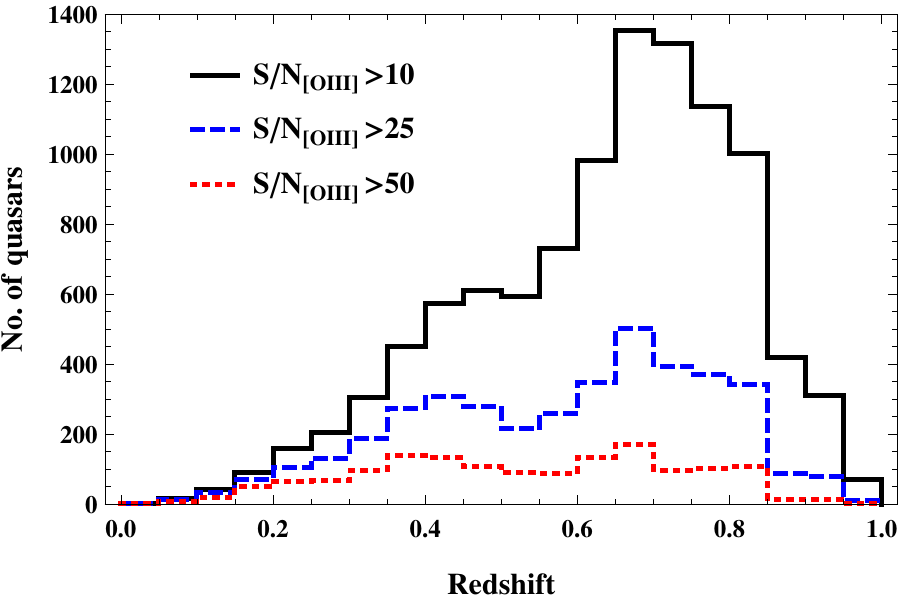}}
\caption{Left-hand panel: sky distribution of the full SDSS-III/BOSS DR12Q quasars (297\,301) in J2000 equatorial coordinates. Right-hand panel: number of quasars with \oxygen emission lines in our fiducial sample (10\,363 quasars) in $\Delta z= 0.05$ bins. S/N$_{\oxygen 5008}>10$ (10\,363 quasars), black solid line; S/N$_{\oxygen 5008}>25$ (4015 quasars), blue dashed line; and S/N$_{\oxygen 5008}>50$ (1498 quasars), red dotted line.}
\label{fig:QSO}
\end{figure*}

The beginning of the SDSS survey opened a new era of precision, allowing us to use big samples of quasars; thus, reducing the statistical uncertainty of the measurement of $\Delta\alpha/\alpha$ (see Table \ref{tab:comparison}).
% \citet{Bahcall} analysed a quasar sample drawn from the SDSS-Early Data Release \citep[SDSS-EDR,][]{EDR}. They obtained a value of 
%
%\begin{eqnarray}
%\frac{\Delta\alpha}{\alpha}= \left(0.7 \pm 1.4\right) \times 10^{-4}\,
%\end{eqnarray}
%
%for a sample of 42 quasars with redshifts $z<0.8$, this period covers the last 7 Gyr. With 1,568 spectra in the same redshift interval, \citet{Gutierrez} reported a more precise value using the SDSS Data Release 6 \citep[SDSS-DR6,][]{DR6}, i.e.
%
%\begin{eqnarray}
%\frac{\Delta\alpha}{\alpha}= \left(2.4 \pm 2.5\right) \times 10^{-5}\,.
%\end{eqnarray}
%
%Recently \citet{Rahmani} used the SDSS-DR7 \citep{DR7} and measured
%
%\begin{eqnarray}
%\frac{\Delta\alpha}{\alpha}= \left(-2.1 \pm 1.6\right) \times 10^{-5}\,
%\end{eqnarray}
%
%for 2,347 quasars, where the quoted error for this measurement is the standard deviation of the results weighted by errors estimated from Monte Carlo simulations.
Here, we extend these works by using the SDSS-III/BOSS Data Relase 12 \citep[SDSS-DR12;][]{DR12}, which covers the full Baryonic Oscillation Spectroscopic Survey (BOSS) survey footprint with an area coverage of 10\,000 deg$^2$. In contrast to these previous investigations, we use spectra obtained with the current BOSS spectrograph \citep{Smee} instead of the previous SDSS-I/II instrument, making our BOSS sample totally independent from previous works.  Moreover, the spectral range of the BOSS spectrograph allows an extension of the redshift interval for the \oxygen doublet from $z=0.8$ to $z=1$. The number of quasar spectra is increased by a factor of 5 with respect to SDSS-DR7. All these spectra have been visually inspected and classified as quasars by the BOSS collaboration, and their products are provided in the SDSS-III/BOSS Data Release 12 Quasar catalogue \citep[DR12Q; see ][]{Paris12}. For the final constraint on $\Delta\alpha/\alpha$, we combine in this work the BOSS sample with the previously studied SDSS-II DR7 quasar sample.

%In this article, we consider the possible variation of the fine-structure constant $\Delta\alpha/\alpha$ as measured through astronomical observations. For this purpose, we use the complete dataset of quasar spectra collected by the Sloan Digital Sky Survey III \citep[SDSS-III,][]{Eisenstein} Baryonic Oscillation Spectroscopic Survey \citep[BOSS,][]{Dawson}. 

There are several emission doublets, in addition to \oxygen (\lines), that can be used to measure $\Delta\alpha/\alpha$ as noted by \citet{Bahcall}, and first used by \citet{Grupe}. \citet{Gutierrez} analysed different doublets and found that the \neon (\linesNe) and \mbox{[Si {\sc ii}]} (\linesSi) doublets appear in quasar spectra with sufficient frequency to have a meaningful sample. Results for [Si {\sc ii}] are consistent with no variation of the fine-structure constant, although the uncertainty is an order of magnitude bigger than for \oxygen, and this doublet can only be used at low redshift $<0.4$ for optical spectra. However, they obtained a positive variation of the fine-structure constant, $\Delta\alpha/\alpha=\left(34\pm1\right)\times 10^{-4}$, when the \neon lines are used. No explanation was found for this positive variation. In this work, we also analyse the \neon lines to check whether the same effect is present in our BOSS quasar sample.

There are investigations which use Si {\sc iv} absorption lines ($\lambda\lambda\ 1394, 1403\ \AA$) to obtain a precision of $4\times10^{-6}$ \citep{Chand}. This method also avoids the assumption of ionization and chemical homogeneity. However, since the separation between both lines is only $\approx 9\ \AA$, the wavelength precision needed in the laboratory for the separation between both lines is five times higher than using \oxygen lines. Nevertheless, these constraints apply to the redshift interval $1.59<z<2.92$, which does not overlap with our range, thus they are complementary to the ones reported in this research.

Finally, in the light of the upcoming large galaxy surveys, like eBOSS and DESI, that will provide millions of high-redshift galaxy spectra, we also discuss using galaxies instead of quasars to set constraints on the fine-structure constant.

The paper is organized as follows. First, in Section \ref{sec:sample}, we describe the data set used for our analysis. Next, in Section \ref{sec:methodology}, the methodology is presented, the emission-line method is explained, and the code and simulations to analyse the spectra are described. In Section \ref{sec:systematics}, we study several samples to check for systematics. Then, our results are presented in Section \ref{sec:results}. Finally, we provide in Section \ref{sec:summary} a summary of the main conclusions achieved with this research project.

%%%%%%%%%%%%%%%%%%%%%%%%%%%%%%%%%%%%%%%%%%%%%%%%%%%%%%
%%%%%%%%%%%%%%%%%%%%%%%%%%%%%%%%%%%%%%%%%%%%%%%%%%%%%%

\section{Sample description}
\label{sec:sample}
%%%%%%%%%%%%%%%%%%%%%%%%%%%%%%%%%%%%%%%%%%%%%%%%%%%%%%
%%%%%%%%%%%%%%%%%%%%%%%%%%%%%%%%%%%%%%%%%%%%%%%%%%%%%%

All the spectra used in this investigation were downloaded from the SDSS Database. This survey \citep{York}, which began taking observations in 1998, consists of a massive collection of optical images and spectra from astronomical objects including stars, galaxies and quasars. For this purpose, there is a dedicated 2.5-m wide-angle optical telescope at Apache Point Observatory in New Mexico \citep[USA; for more details, see][]{Gunn}. The third phase of this project \citep[SDSS-III;][]{Eisenstein} includes BOSS \citep{Dawson} among its four main surveys. The data analysed in this research were provided by BOSS, and it is used for measuring $\Delta\alpha/\alpha$ for the first time. The SDSS-III/BOSS pipeline \citep{Bolton} classifies the objects as quasars with a $\chi^2$ minimization procedure to fit the observed spectrum to multiple  galaxy and quasar spectrum templates for all allowed redshifts. Then, a visually-inspected quasar catalogue is built from these objects. Our fiducial sample is obtained from the DR12Q catalogue version \citep{Paris12}.

The wavelength coverage of the SDSS-III/BOSS spectrograph is 3600-10\,400\ \AA\ and that of the SDSS-II spectrograph is 3800-9200 \AA.
The BOSS sample is homogeneous since all the spectra have been obtained with the same instrument, and it is independent from previous investigations. The wider coverage of the new spectra allows consideration of higher redshifts (up to $z=1$ for \oxygen doublet) than in the previous SDSS-II analysis based on the same method (see Table \ref{tab:comparison}). The BOSS spectrograph has two channels (blue and red) whose wavelength coverage is $3600$-$6350\,\AA$ and $5650$-$10400\,\AA$, respectively. The resolving power ranges from 1560 at $3700\,\AA$ to 2270 at $6000\,\AA$ (blue channel) and from 1850 at $6000\,\AA$ to 2650 at $9000\,\AA$ (red channel). For our sample, the \oxygen lines fall in the red channel for $>96\,\%$ of the quasars. The number of pixels of each spectrum is about $4600$ for the BOSS spectra and $3800$ for the SDSS-I/II spectra. The pixel spacing is uniform in log-wavelengths ($\Delta\log\lambda=10^{-4}\, \rm{dex}$). More complete information about the SDSS-I/II and BOSS spectrographs can be found in \citet{Smee}.

\subsection{Data selection}
\label{sec:sample1}
\begin{figure*}
\centerline{\includegraphics[angle=0,width=0.495\textwidth]{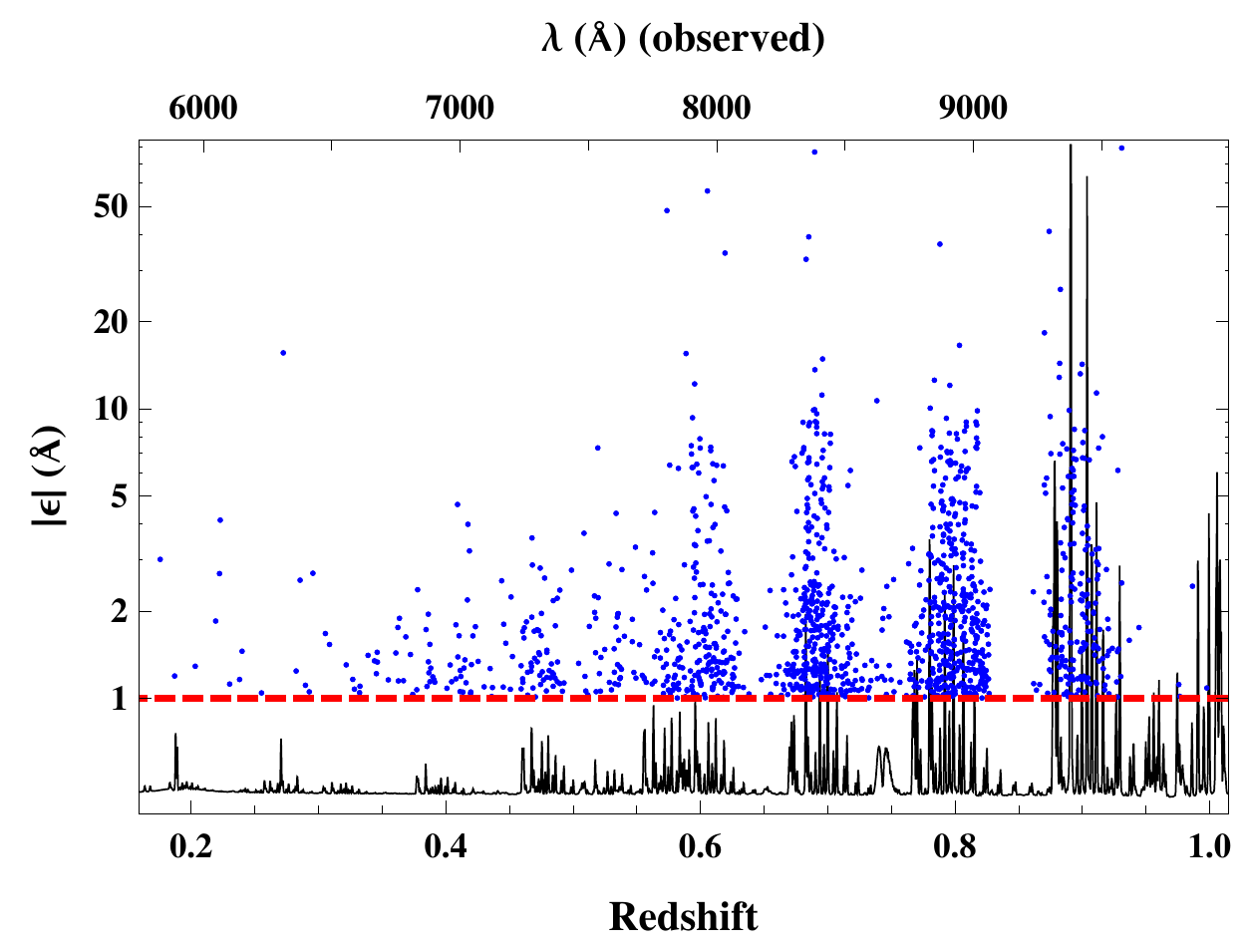}\hfill \includegraphics[angle=0,width=0.495\textwidth]{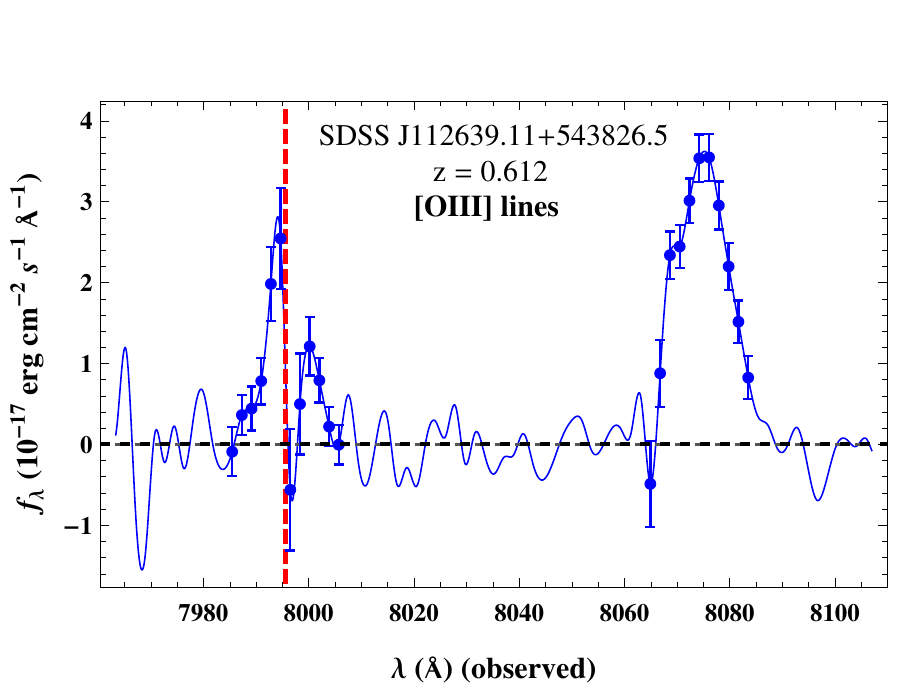}}
\caption{Left-hand panel: data points (1416) for which $|\epsilon|=|\delta\lambda_{z}/(1+z)-\delta\lambda_{0}|$, namely the absolute value of the difference between the measured line separation at redshift $z$ in rest frame and the local one, is bigger than 1\,\AA\ plotted as a function of redshift (and the wavelength observed for \oxygen 4960). We compare with a typical sky spectrum: the \oxygen positions for these spectra correlate with the sky emission lines. Hence, these high values of $|\epsilon|$ are due to bad sky subtractions and/or low S/N. These spectra are removed. Right-hand panel: a spectrum removed from the sample because of the sky emission-line criteria. For this quasar, we get $\epsilon=1.2\pm0.6$\,\AA%$\Delta\alpha/\alpha=\left(12 \pm 6\right)\times 10^{-3}$
. The weak \oxygen line is affected by the subtraction of the 7995 \AA\ OH sky emission line, indicated by the vertical red dashed line.}
\label{fig:skycontamination}
\end{figure*}

The SDSS-III/BOSS DR12Q catalogue contains 297\,301 objects. Fig.\ \ref{fig:QSO} (left-hand panel) shows the quasar distribution in the sky. We summarize below the main selection criteria in order to define our fiducial sample from this catalogue.

\begin{enumerate}

\item{Redshift $<1$}. This limitation is imposed by the wavelength range of the BOSS optical spectrograph and the position of the \oxygen lines. This criterion decreases the sample down to 45\,802 quasars. \smallskip

\item{S/N$_{\rm{\oxygen 5008}}>10$.} We impose a mild constraint on the signal-to-noise ratio (S/N) of the stronger \oxygen line (5008 \AA) in order to preserve a large number of spectra. Constraints on the expected width and amplitudes of the lines help in avoiding misidentifications of the \oxygen doublet (see Section \ref{sec:systematics}). This selection reduces the sample from 45\,802 to 13\,023 objects.\smallskip

\item{Non-converging fits.} Since we analyse spectra with low S/N, there are some cases where the Gaussian fit to the lines does not converge. 1\,244 spectra are discarded, leaving us with 11\,779 spectra.\smallskip

\item{Sky emission lines.} Strong atmospheric lines, for instance the \mbox{O {\sc i}} 5578 \AA\ line, are poorly or not completely removed by the SDSS sky subtraction algorithm. This may lead to a wrong identification of the \oxygen lines and to include low S/N$_{\rm\oxygen}$ spectra \citep{Gutierrez}. Both effects will produce outliers.  We use the SDSS sky mask for Ly$\alpha$ forest studies which contains 872 lines \citep[see][for more details]{Delubac} to remove spectra whose \oxygen lines lie within a particular distance from the strongest sky lines. Even though we vary the distance \oxygen -- sky lines, use different set of sky lines (according to their intensity), or evaluate other conditions (S/N, fit errors, etc.) to remove affected spectra; we usually eliminate $3-5$ good spectra for each bad spectra eliminated. Thus, these tests decrease significantly the number of quasars while not being very effective: typically $50\%$ of the outliers are not removed. Thus, we decided to eliminate all spectra for which the separation between both lines differ by more than 1 \AA\ from the local value (see the last paragraph in \mbox{Section \ref{sec:methodology3}}).  Fig.\ \ref{fig:skycontamination} (left-hand panel) shows that the distribution of these outliers is correlated with a typical sky spectrum. From a visual inspection, we observed that these spectra have low S/N, and they are in fact contaminated by sky emission line subtraction (see right-hand panel of Fig.\ \ref{fig:skycontamination}). This effect causes us to discard 1\,416 spectra ($12\%$ of the previous 11\,779 quasars). Finally, we have 10\,363 quasar spectra (our `fiducial sample').

\end{enumerate}

The presence of broad H\,$\beta$ emission line (4861 \AA) near the weak \oxygen line 4960 \AA\ could produce a blueshift in the determination of the \oxygen line position. This could mimic a positive variation on the fine-structure constant. Therefore, a constraint on the strength and/or width of the H\,$\beta$ emission line has been imposed on previous investigations \citep{Bahcall,Gutierrez,Rahmani}. However, we do not restrict any characteristic of the H\,$\beta$ line in our fiducial sample. We obtain a weighted mean for $\Delta\alpha/\alpha$ using as weights the uncertainty in $\Delta\alpha/\alpha$ computed with the standard errors for the position of the lines derived from the Gaussian fits. The contamination of H\,$\beta$ is automatically taken into account. For instance, a broad H\,$\beta$ line near the \oxygen 4960 line means a bad Gaussian fit. Thus, we obtain larger errors in the position of the line centroids and, consequently, in $\Delta\alpha/\alpha$. In Section \ref{sec:systematics}, we analyse several samples where the S/N$_{\rm{H}\,\beta}$ is constrained to check that the H\,$\beta$ contamination has little weight on the final constraint value.

An electronic table is published along with the paper which contains all the information of each spectrum from our fiducial sample of 10\,363 quasars (see Appendix A). 

\begin{figure*}
\centerline{
\includegraphics[angle=0,width=0.52\textwidth]{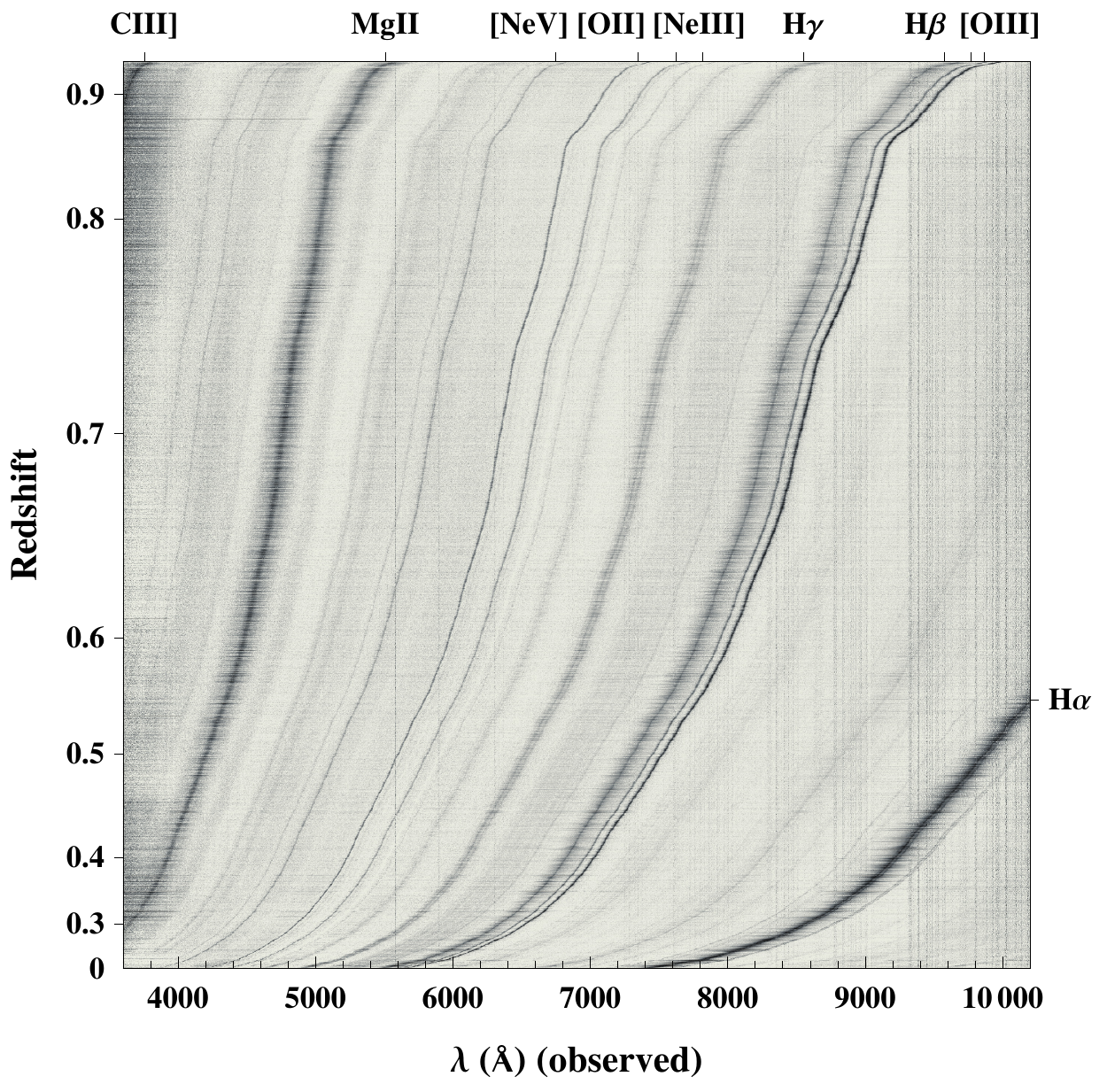}\hspace{-0.6cm}%\hspace{-0.6cm}
 \includegraphics[angle=0,width=0.52\textwidth]{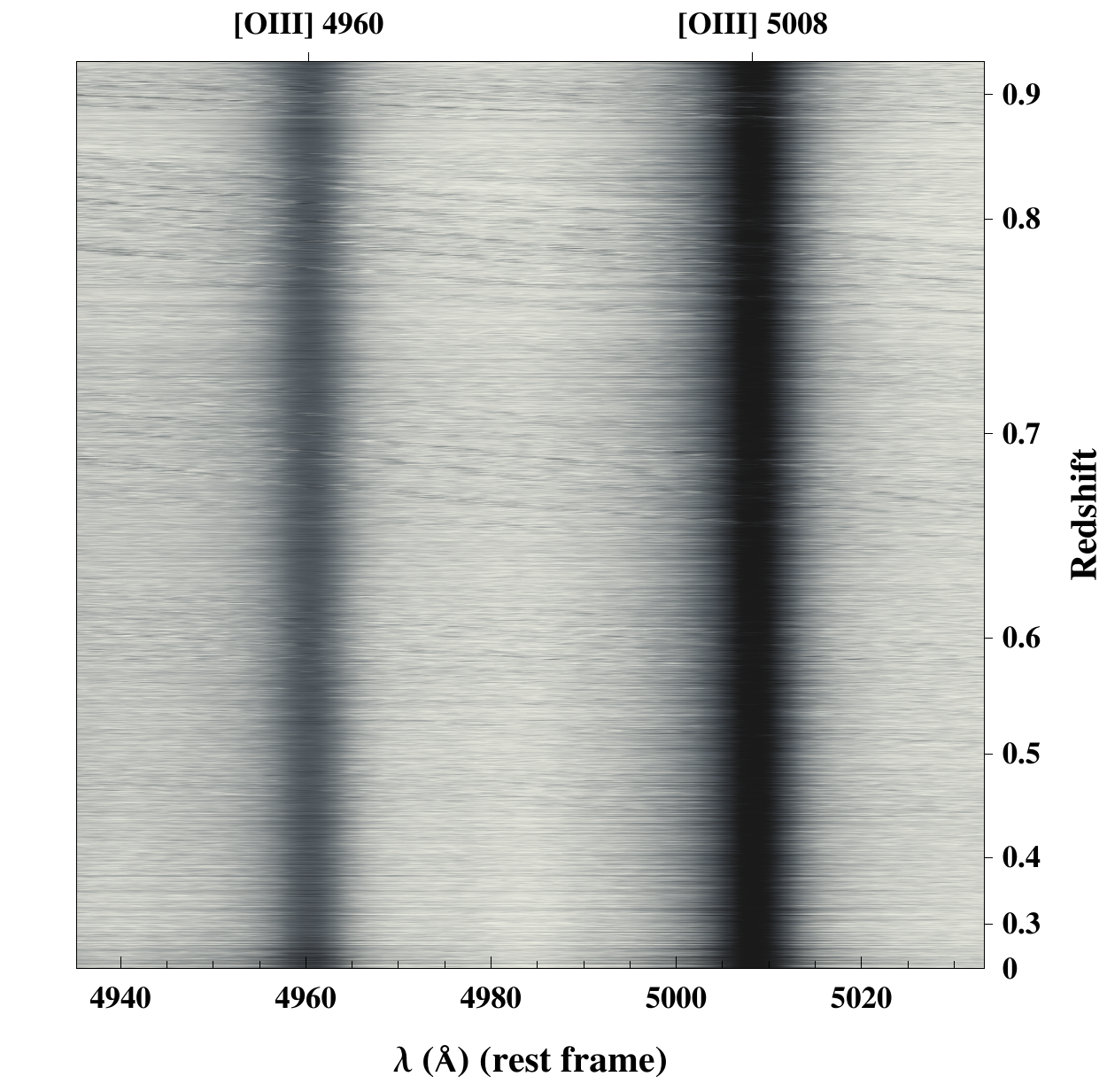}}
\caption{Composite image with our fiducial sample of 10\,363 BOSS quasar spectra sorted by redshift. Left-hand panel: the whole range of wavelengths is shown. From right to left, the strongest emission lines are H\,$\alpha$ 6565 \AA; \oxygen \lines; H\,$\beta$ 4861 \AA; H\,$\gamma$ 4341 \AA; \neon \linesNe; [O {\sc ii}] 3730 \AA; [Ne {\sc v}] 3426 \AA; Mg {\sc ii} 2796 \AA\ and C {\sc iii}] 1906 \AA. The narrow straight line at 5579 \AA\ is the strong [O {\sc i}] atmospheric line. Right-hand panel: wavelength interval centred at the \oxygen doublet in rest frame.}
\label{fig:stack}
\end{figure*}

The distribution of the selected quasars in redshift according to their selected S/N$_{\oxygen 5008}$ is plotted in Fig.\ \ref{fig:QSO} (right-hand panel). Fig.\ \ref{fig:stack} (left-hand panel) displays a composite image built with all the spectra from our fiducial sample sorted by redshift. The right-hand panel shows the \oxygen doublet in rest frame.

%%%%%%%%%%%%%%%%%%%%%%%%%%%%%%%%%%%%%%%%%%%%%%%%%%%%%%
%%%%%%%%%%%%%%%%%%%%%%%%%%%%%%%%%%%%%%%%%%%%%%%%%%%%%%

\section{Methodology}
\label{sec:methodology}
%%%%%%%%%%%%%%%%%%%%%%%%%%%%%%%%%%%%%%%%%%%%%%%%%%%%%%
%%%%%%%%%%%%%%%%%%%%%%%%%%%%%%%%%%%%%%%%%%%%%%%%%%%%%%

\subsection{Measurement method}

To first order, the difference between the energy levels of an atom is proportional to $\alpha^2$. Transitions between energy levels of the same atom at a given ionization level, with the same principal quantum number and different total angular momentum $J$, have an energy difference proportional to $\alpha^4$. These groups of transitions are called fine-structure multiplets. \cite{Savedoff} first realized that the fine structure of these energy levels could be used to break the degeneracy between the redshift effect and a possible variation of $\alpha$. 

The value of the fine-structure constant can be measured through the separation between absorption or emission multiplets in the spectra of distant quasars \citep{Uzan2003} as
\begin{eqnarray}
\frac{\Delta\alpha}{\alpha}\left(z\right)\equiv\frac{1}{2}\left\{ \frac{\left[\left(\lambda_{2}-\lambda_{1}\right)/\left(\lambda_{2}+\lambda_{1}\right)\right]_{z}}{\left[\left(\lambda_{2}-\lambda_{1}\right)/\left(\lambda_{2}+\lambda_{1}\right)\right]_{0}} -1 \right\}\,,
\label{alpha}
\end{eqnarray}
where $\lambda_{1,2}\, (\lambda_2>\lambda_1)$ are the wavelengths of the transitions and subscript 0 and $z$ stand for their value at redshift zero (theoretical/laboratory values) and at redshift $z$, respectively.
For illustrative purposes, expression (\ref{alpha}) can be approximated by
\begin{eqnarray}
\frac{\Delta\alpha}{\alpha}\approx \frac{\epsilon}{2\,\delta\lambda_{0}}\,,
\label{formula}
\end{eqnarray}
where $\delta\lambda_{0}=\left[\lambda_2-\lambda_1\right]_0$ is the local $z=0$ separation between both wavelengths, and $\epsilon=\delta\lambda_{z}/(1+z)-\delta\lambda_{0}$ is the difference between the measured line separation at redshift $z$ in rest frame and the local one. Thus, in principle, the larger the difference between the pair of lines, the better the precision for measuring $\Delta\alpha/\alpha$. %However, in practice, systematics in the wavelength calibration over a larger interval dominate.

Concerning emission lines, the most suitable pair of lines is the \oxygen doublet, which is often present in quasar spectra with relatively high-S/N. The vacuum values for the \oxygen doublet wavelengths are
\begin{eqnarray}
\lambda^{\rm{\oxygen}}_{1}=4960.295 \;\rm{\AA}\, \hspace{0.6cm} \lambda^{\rm{\oxygen}}_{2}=5008.240\;\rm{\AA}\,
\end{eqnarray}
\begin{eqnarray}
\delta\lambda^{\rm{\oxygen}}_{0}= 47.945 \;\rm{\AA}\,,
\label{sep}
\end{eqnarray}
which are published in the NIST Atomic Spectra Database.\footnote{\href{http://physics.nist.gov/PhysRefData/ASD/lines\_form.html}{http://physics.nist.gov/PhysRefData/ASD/lines\_form.html}} %These values correspond to heliocentric vacuum wavelengths since the SDSS wavelength calibration is based on vacuum wavelengths.
These transitions are forbidden (they correspond to magnetic dipole and electric quadrupole transitions), and they are not observed in the laboratory. The wavelength experimental values are obtained indirectly by first computing the energy levels from observed wavelengths using a theta-pinch discharge \citep{Pettersson}. The wavelength separation has directly been measured in the infrared from H {\sc ii} regions using a balloon-borne telescope and Michelson interferometer \citep{Moorwood}. Both measurements of the wavelength separation, indirectly with the theta-pinch discharge and directly with the Michelson interferometer,  are in good agreement, being the Michelson interferometer more accurate with an error $<5\times 10^{-4}$ \AA.

From equation (\ref{formula}), a determination of $\epsilon$ with a precision of $1\, \AA$ allows for an uncertainty of $10^{-2}$ in $\Delta\alpha/\alpha$ when using the \oxygen doublet. The precision from the NIST atomic data allows for a determination of $\Delta\alpha/\alpha$ up to $10^{-5}$, which is a bit less than the uncertainty in our result. One could perform a blind analysis in order to search for a possible variation on $\alpha$, where the absolute wavelength values are not required, if one had a large enough sample distributed in redshift. However, the precision on the absolute wavelengths limits the usefulness of high-resolution spectroscopy until better measurements of the \oxygen lines (or just their separation) are available.

\subsection{Implementation}

\begin{figure*}
\centerline{\includegraphics[angle=0,width=0.36\textwidth]{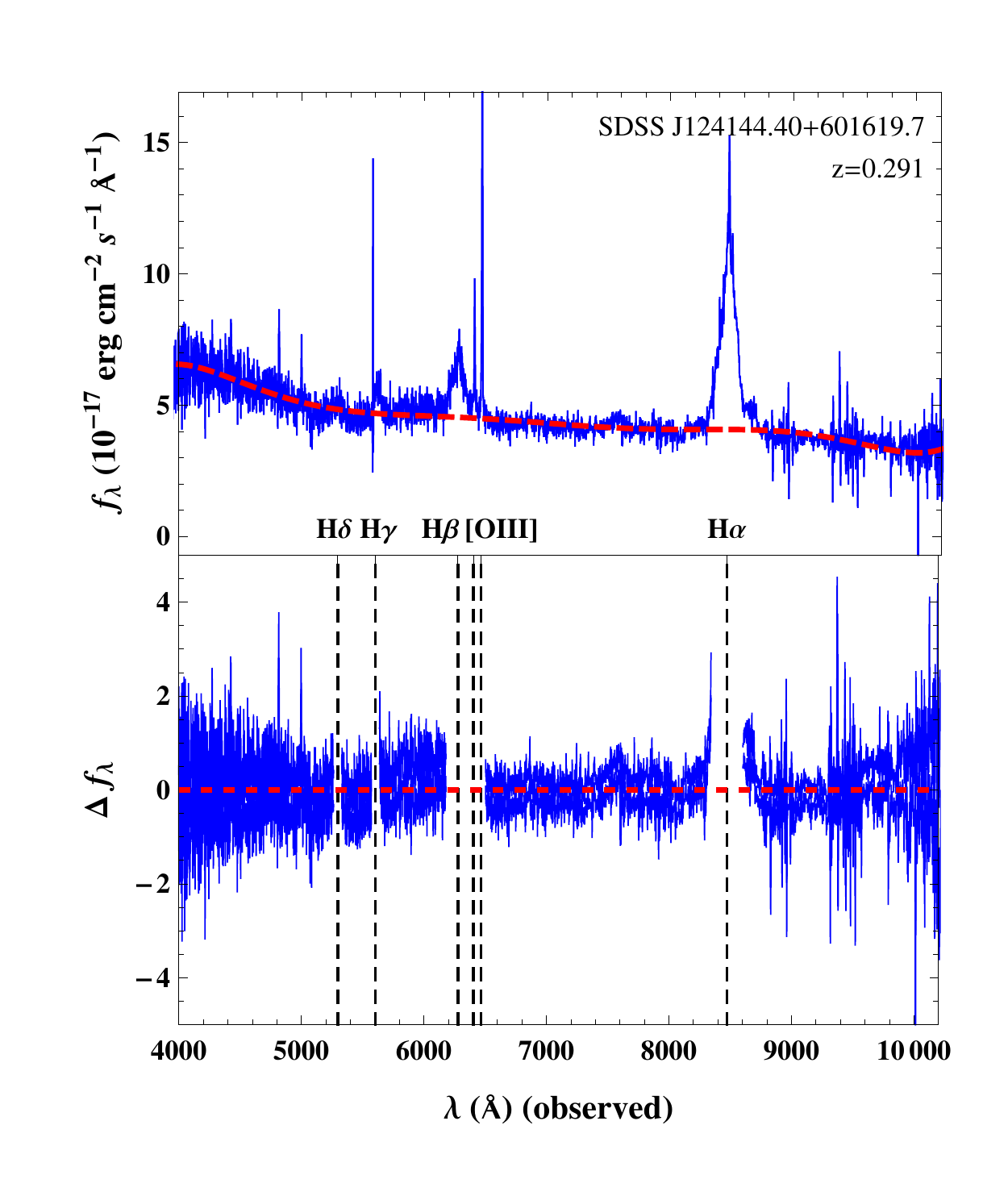}\hspace{-0.5cm}\includegraphics[angle=0,width=0.36\textwidth]{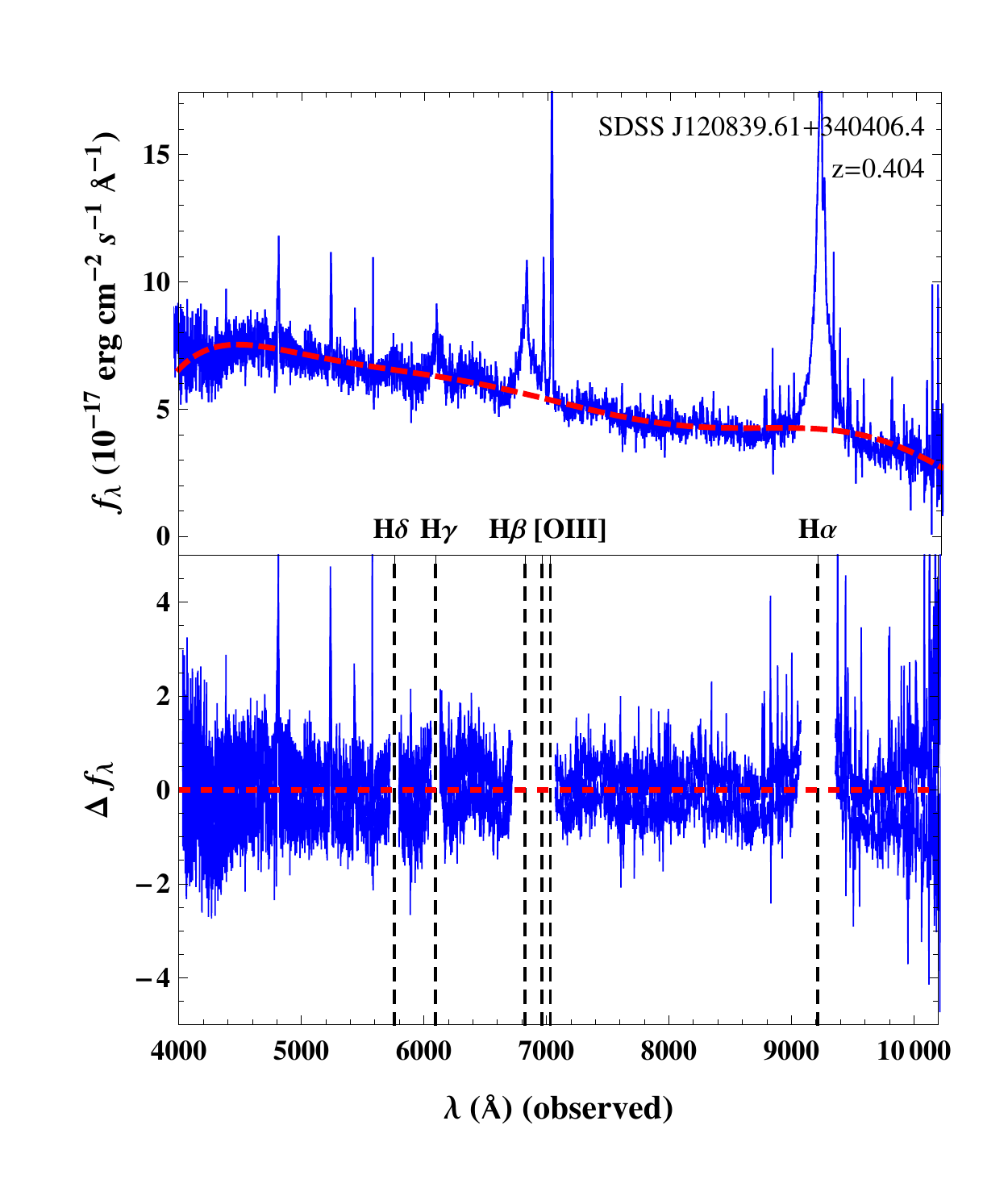}\hspace{-0.5cm}
			\includegraphics[angle=0,width=0.36\textwidth]{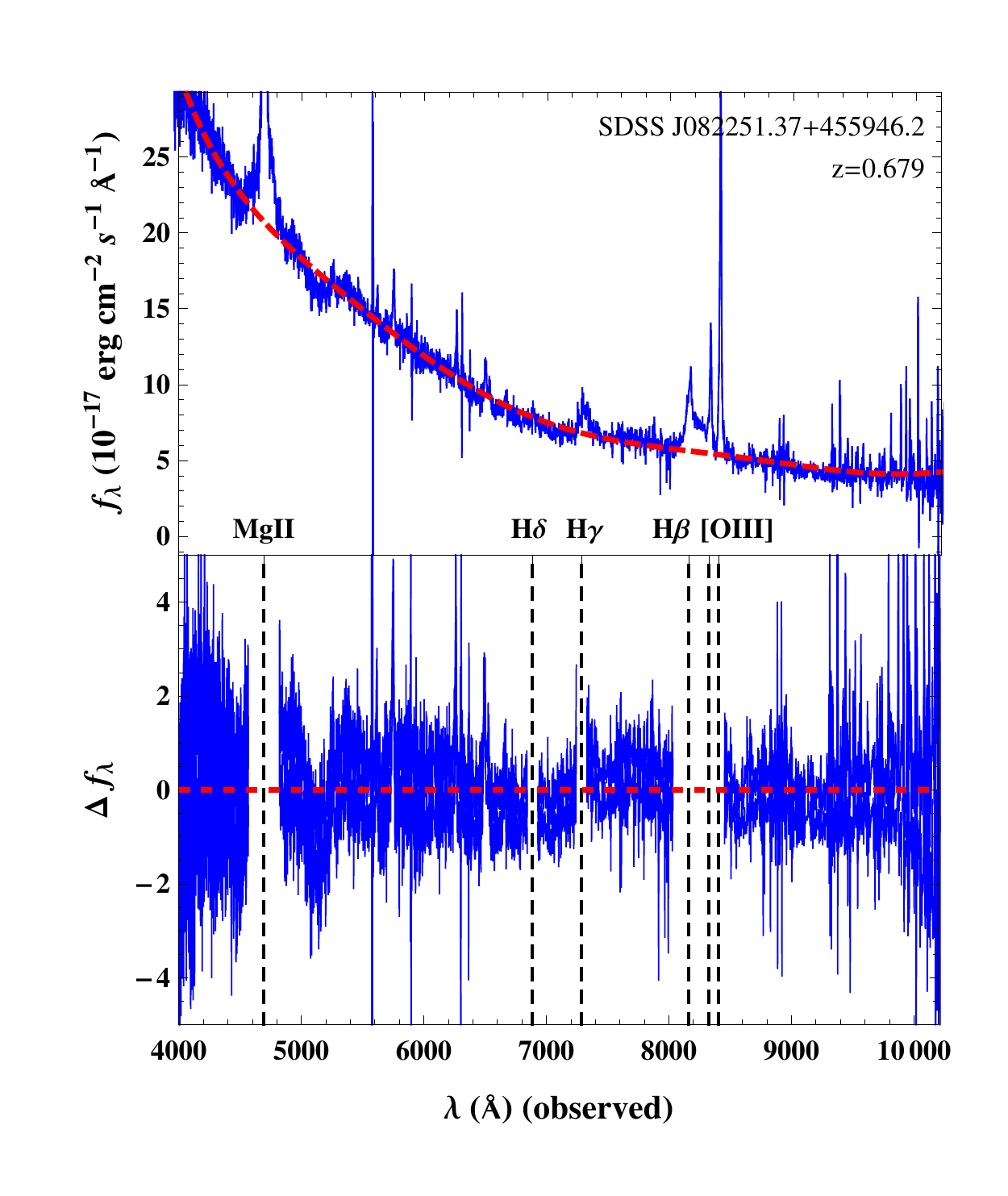}}
			\vspace{-0.5cm}
\caption{Seventh-order polynomial fits (red) to the continuum spectrum with their residuals for three typical quasar spectra at different redshift. The gaps in the residuals are the masked regions corresponding to (from right to left) $\rm{H}\,\alpha$, the \oxygen doublet, $\rm{H}\,\beta$, $\rm{H}\,\gamma$, $\rm{H}\,\delta$ and Mg {\sc ii}  (black dashed lines).
% In Section \ref{sec:systematics}, we study whether the chosen polynomial order for the continuum fits affects our measurement of $\Delta\alpha/\alpha$.
}
\label{fig:continuum}
\end{figure*}

The code developed for the analysis of the quasar spectra follows the one described in \citet{Gutierrez}, although there are some modifications and more information has been extracted from the analysis. We describe the main characteristics of our code below.
\vspace{-0.5cm}
\subsubsection{Wavelength sampling}
We consider only the experimental data together with their errors as processed by the SDSS pipeline to obtain the constraint on the possible variation of $\alpha$. We do not resample the wavelength range by using an interpolation method. Since the pixel spacing is uniform in log-wavelengths, a given range of wavelengths in rest frame $(\lambda_{-},\lambda_{+})$ has the same number of pixels $N$, i.e.
\begin{eqnarray}
N \propto \int^{\lambda_{+}\,(1+z)}_{\lambda_{-}\,(1+z)} \rm{d}\left(\log{\lambda}\right)=\log{\frac{\lambda_{+}\,(1+z)}{\lambda_{-}\,(1+z)}}=\log{\frac{\lambda_{+}}{\lambda_{-}}}\,,
\end{eqnarray}
and is independent of the redshift of the object. All the wavelength intervals with the same width in rest frame will have the same number of experimental points.

\vspace{-0.5cm}
\subsubsection{Fit of the continuum spectrum}

First, we fit a seventh-order polynomial to subtract the continuum spectrum while masking regions where strong and wide emission lines are present ($\rm{H}\,\alpha$, $\rm{H}\,\beta$, $\rm{H}\,\gamma$, $\rm{H}\,\delta$, Mg {\sc ii} and the \oxygen doublet). Our method differs from \citet{Gutierrez} in that they use a cubic local spline to fit the continuum masking strong emission lines. The chosen order of the polynomial provides enough degrees of freedom to reproduce different continuum features. In Section \ref{sec:methodology}, we test how our measurement for $\Delta\alpha/\alpha$ is affected by changing the polynomial order. Hundreds of continuum spectra fits were checked by eye. The residuals from the fits are smaller than the errors on the flux densities. Fig.\ \ref{fig:continuum} shows three different spectra with their continuum fit and residuals. 

\vspace{-0.5cm}
\subsubsection{Signal-to-noise ratio}

We follow \citet{Gutierrez} for the determination of S/N. Hence, we compute the standard deviation of the flux between $5040\,(1+z)$ and $5100\,(1+z)\,\AA$ (where $z$ is the redshift of the quasar) where there are no strong emission or absorption lines. Then, we search for the maximum of the \oxygen 5008 line, and determine S/N$_{\rm{\oxygen 5008}}$ as the ratio between the maximum of the line and the previously computed standard deviation. Although for a more reliable determination of the S/N, it is better to use a Gaussian fit to the line. This procedure avoids possible issues related when fitting data with very low S/N. This S/N is used in the criterion ii (Section \ref{sec:sample}) to build our fiducial sample.

\vspace{-0.5cm}
\subsubsection{Measurement of the emission-line wavelengths}
\label{sec:methodology24}

To measure the wavelengths of the \oxygen doublet, our fitting code needs as input an accurate estimate of the redshift of the quasar, at least with an error \mbox{$\Delta z<3\times10^{-3}$}. This allows a search for the emission lines in a $15\,\AA$ window around the expected location of the \oxygen lines. The SDSS pipeline provides a determination of the redshift based on a $\chi^2$ fit to different templates; we refer to \citet{Bolton} for more details. These redshift estimates have errors between $10^{-4}$ and $10^{-5}$, which are sufficient for our purposes. Moreover, there is also a visual redshift estimation which can be found in the quasar catalogue DR12Q \citep{Paris12}. The difference between both redshift estimates (if any) is usually $|z_{\rm vis}-z_{\rm pipe}|\approx5\times10^{-4}$. We decided to adopt the visual redshifts.

\begin{figure*}
\centerline{\includegraphics[angle=0,width=0.5\textwidth]{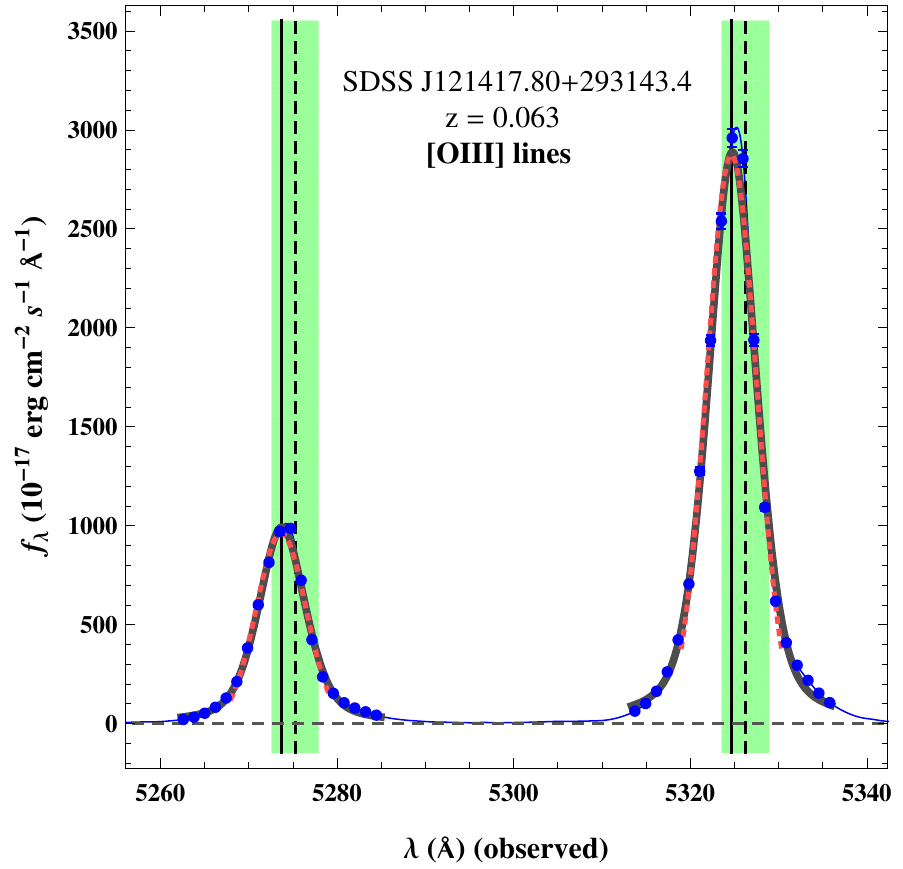}\hfill
			\includegraphics[angle=0,width=0.5\textwidth]{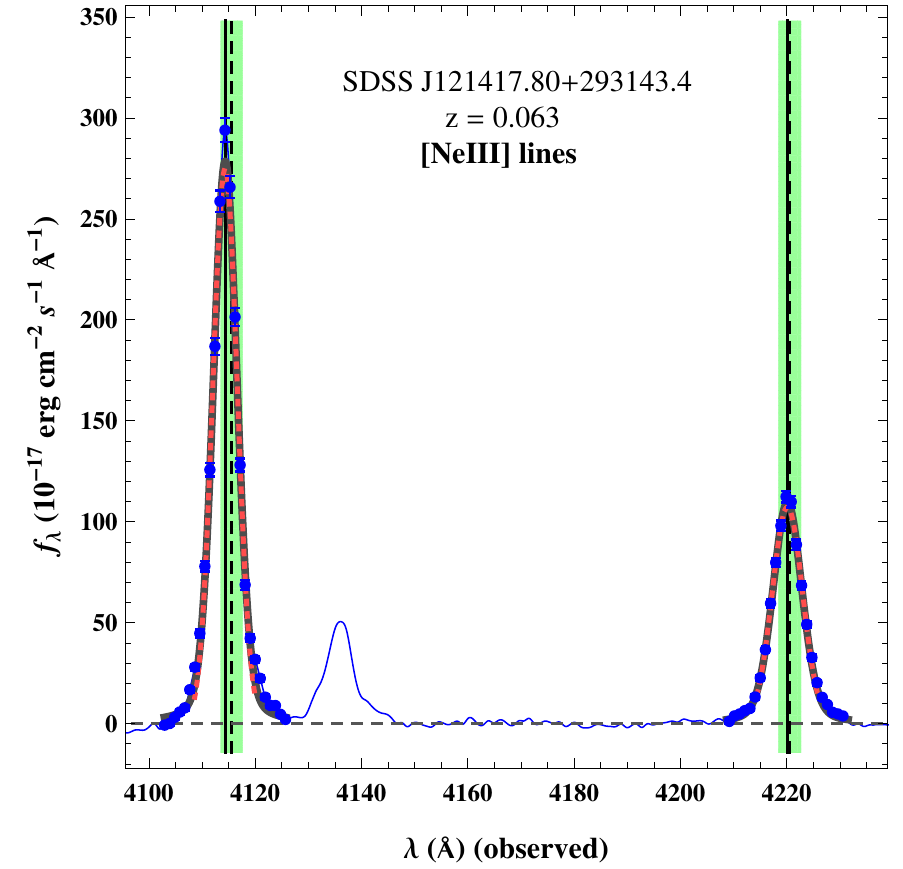}}
\caption{\oxygen (left-hand panel) and \neon (right-hand panel) lines for SDSS-J121417.80+293143.4, at redshift $z=0.063$. The measured $\Delta\alpha/\alpha$ for this quasar is $\Delta\alpha/\alpha^{\rm{Gauss}}_{\rm{\oxygen}}=\left(2.3\pm7.6\right)\times10^{-4}$, $\Delta\alpha/\alpha^{\rm{Voigt}}_{\rm{\oxygen}}=\left(3.3\pm12.6\right)\times10^{-4}$ and $\Delta\alpha/\alpha^{\rm{Gauss}}_{\rm{\neon}}=\left(39\pm8\right)\times10^{-4}$, $\Delta\alpha/\alpha^{\rm{Voigt}}_{\rm{\neon}}=\left(37\pm9\right)\times10^{-4}$. The measured $\Delta\alpha/\alpha$ for \neon is not consistent with zero regardless of the profile; see Fig.\ \ref{fig:NeO} and last paragraph of Section \ref{sec:systematics} for discussion. Each panel shows the flux density for each pixel with their respective error bars (solid symbols), together with the Gaussian fit (dotted red curve) and the pseudo-Voigt profile (thick grey curve) to each of the lines. The fitting procedure (described in the text) only takes into account the experimental data (solid symbols) weighted by their error bars. Notice how the deviation of the line centroid position derived from our Gaussian fit (vertical solid line) with respect to the expected position of the line (vertical dashed line) according to the visual redshift provided by the DR12Q catalogue are well correlated for the same pair of \oxygen and \neon lines, and for the different set of lines. The green shaded vertical areas highlight the uncertainty for the expected position of the lines due to the quasar redshift error ($\approx\, 5\times10^{-4}$). Also shown is a fourth-order spline interpolation to the spectrum after subtracting the continuum (thin solid line). The \neon lines are weaker by one order of magnitude than the \oxygen lines, which is usually the case for all the spectra showing both pair of lines. The weak line near the stronger \neon line is blended with He {\sc i} $(3889.75 \,\AA)$ and H\,$\zeta\ (3890.16 \,\AA)$.}
\label{fig:lines}
\end{figure*}

The centroid positions of the \oxygen emission lines are determined by four different methods.
\vspace{-0.3cm}
\begin{enumerate}

\item{Gaussian profile method.}

First, we search for the maximum flux value in an $\sim15\, (1+z)\, \AA$ window around the expected position of the line (according to the redshift provided by the DR12Q catalogue). This procedure automatically erases any bias produced by the redshift value. Then, we make an initial Gaussian fit around the position of the maximum flux value using a fixed width of $\sim10\,(1+z)\, \AA$. From this first fit, we obtain a new position for the line centroid and a Gaussian width. These values are used as initial parameters for the final fit of the lines; namely, the wavelength range considered to perform the final fit is centred around the position of the line centroid, and it is four times the Gaussian width of the lines. This approach means that we consider pixels up to $2\sigma$ away from the centre of the line. Hence, some lines are fitted using $\sim4-5$ pixels, while others with $\sim 15-20$ pixels depending on the line width. The fit takes into account the flux errors for each pixel, i.e., we use the ${\it ivar}$ column found in each spectrum as weights for the fit. Our final centroid measurement for each considered line corresponds to the centroid of the Gaussian fit done in the last step of the adopted procedure. We also derive an error for $\Delta\alpha/\alpha$ using the standard errors for the centre position of the Gaussians. This is our main method for measuring $\alpha$.

\item{Voigt profile method.}

Following the same procedure than when using a Gaussian profile, we make the fit with a Voigt profile instead of a Gaussian. More precisely, we use a pseudo-Voigt profile which is a linear combination of a Gaussian and a Lorentzian profile. Then, we have one more parameter, i.e.\ the amplitude of the Lorentzian function, while its width and its position are the same as those for the Gaussian \mbox{profile.}

In Fig.\ \ref{fig:lines}, we depict the \oxygen and \neon lines for the same quasar spectrum to illustrate the Gaussian and Voigt fitting \mbox{methods.}

\item{Integration method.}

Here, the centroids of the lines are obtained by integrating around $1\sigma$ from the position of the fitted Gaussian profile. %For narrow lines there are about only $1-2$ pixels to compute the centroid. Hence, we perform a fourth order spline interpolation of the experimental points which is used to perform the integration. This interpolation rises serious concerns but this method is only for comparison and its results on the variation of $\alpha$ must not be taken seriously. 
This technique provides indications of whether there is $\rm{H}\,\beta$ contamination. However, due to the mid-resolution of the spectra $R\approx2000$, this method is not very accurate.

\item{Modified Bahcall method.}

In \citet{Bahcall} the authors used a different approach to compute the line positions. They performed a third-order spline interpolation to the stronger \oxygen 5008 line, then fitted this interpolation to the weaker 4960 line by adjusting the amplitude and separation of the profile. We have modified this method by using a Gaussian fit to the stronger line rather than a third-order spline.

\end{enumerate}

Although we have described four different methods, the main results for $\Delta\alpha/\alpha$ presented in this work are based on the Gaussian fitting method, while the other three are used only for comparison (see Section \ref{sec:systematics}).

Finally, our final result for $\Delta\alpha/\alpha$ and its error is obtained in the same way as in \citet{Chand}, namely we compute a weighted mean and a weighted standard deviation, where the errors for $\Delta\alpha/\alpha$ of each spectrum are used as weights.

%%%%%%%%%%%%%%%%%%%%%%%%%%%%%%%%%%%%%%%%%%%%%%%%%%%%%%%%%%%%%%%%%%%%%%%%%%%%%%%%%%%%%%%%%%%%%%%%%%%%%%%%%%%%
%%%%%%%%%%%%%%%%%%%%%%%%%%%%%%%%%%%%%%%%%%%%%%%%%%%%%%%%%%%%%%%%%%%%%%%%%%%%%%%%%%%%%%%%%%%%%%%%%%%%%%%%%%%%

\subsection{Simulated spectra}
\label{sec:methodology3}
%%%%%%%%%%%%%%%%%%%%%%%%%%%%%%%%%%%%%%%%%%%%%%%%%%%%%%%%%%%%%%%%%%%%%%%%%%%%%%%%%%%%%%%%%%%%%%%%%%%%%%%%%%%%
%%%%%%%%%%%%%%%%%%%%%%%%%%%%%%%%%%%%%%%%%%%%%%%%%%%%%%%%%%%%%%%%%%%%%%%%%%%%%%%%%%%%%%%%%%%%%%%%%%%%%%%%%%%%

\begin{figure*}
\centerline{\includegraphics[angle=0,width=0.5\textwidth]{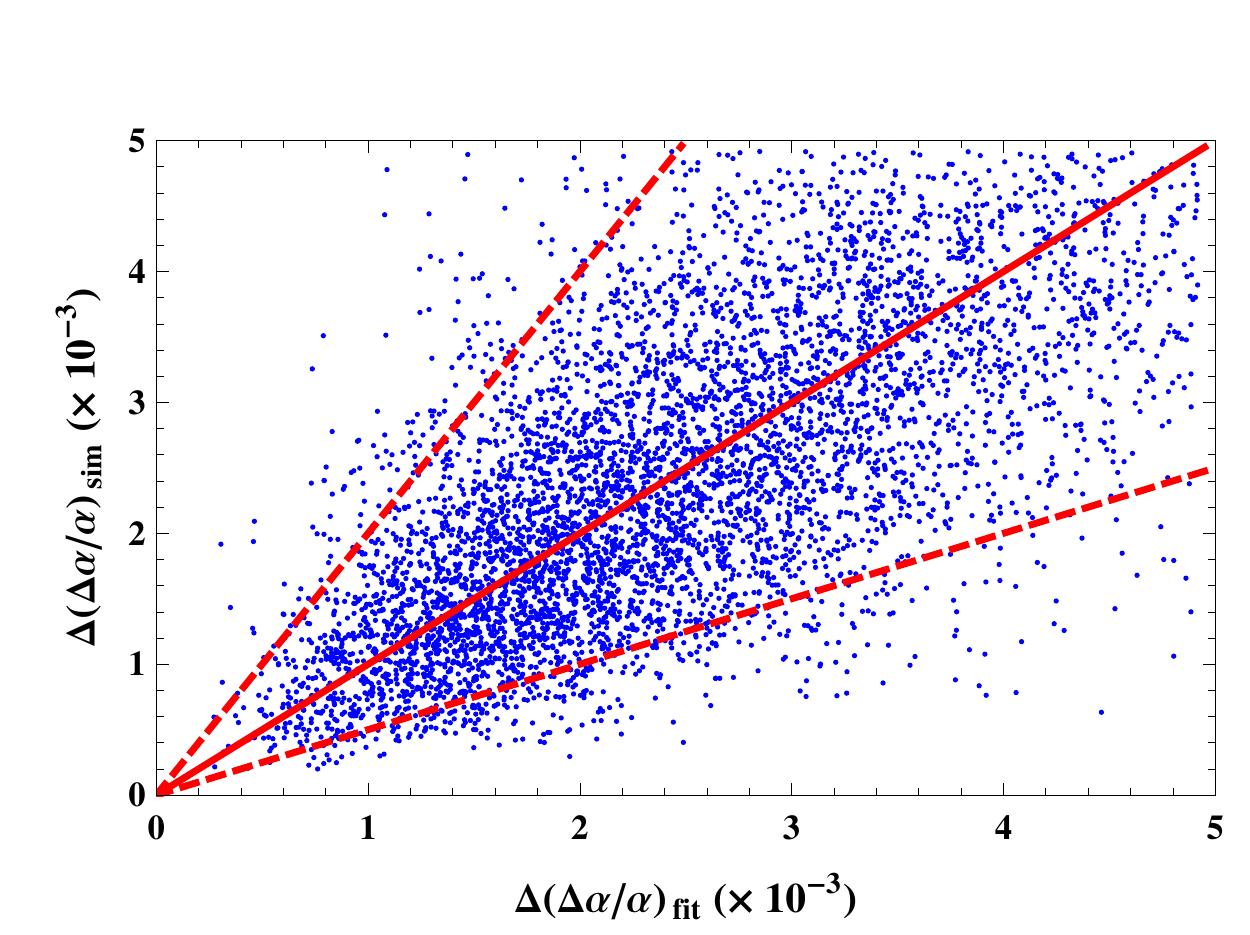}\hfill\includegraphics[angle=0,width=0.5\textwidth]{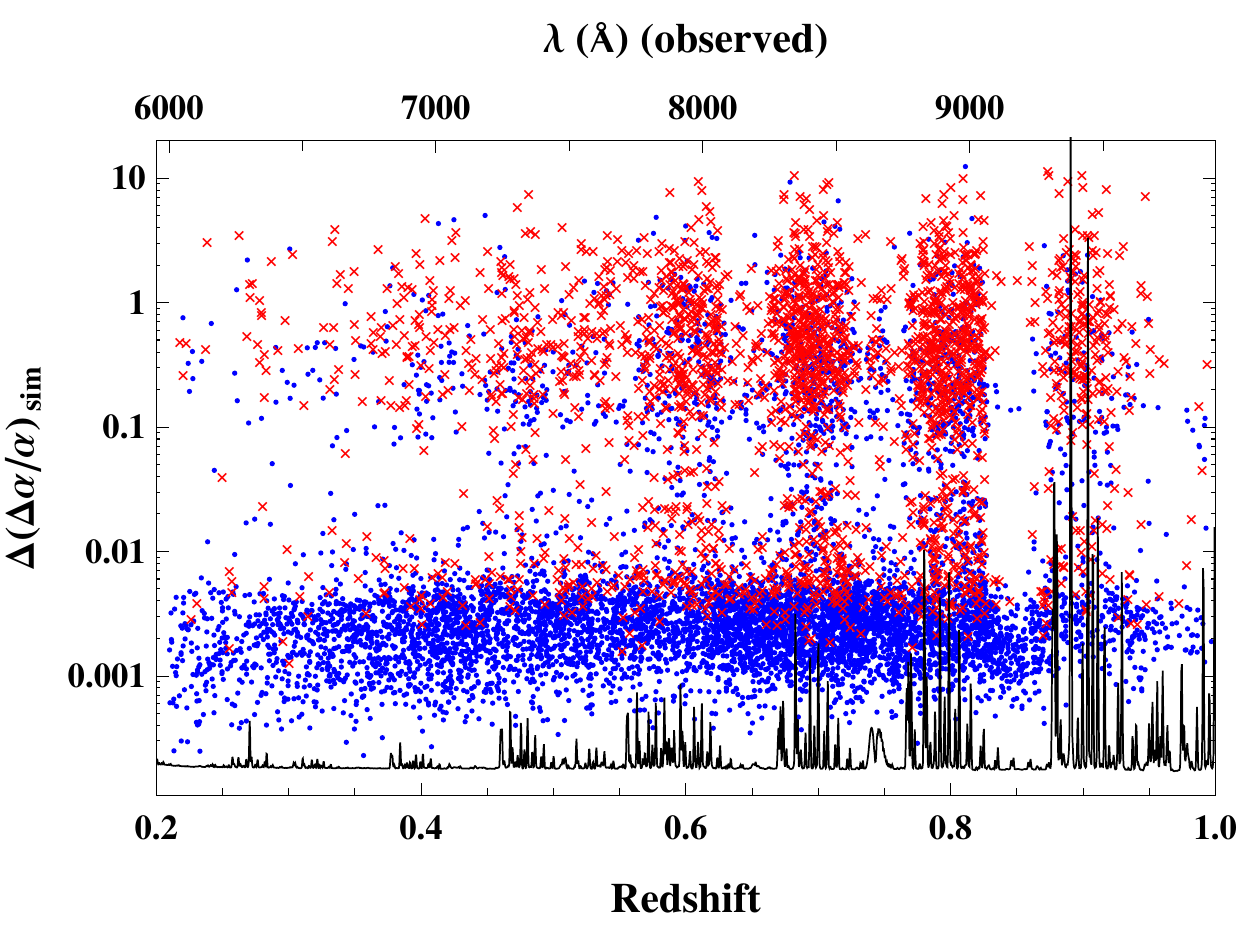}}
\caption{Left-hand panel: errors for $\Delta\alpha/\alpha$ obtained from the simulations (standard deviation of the $\Delta\alpha/\alpha$ measurements on 100 realizations of each real spectrum) and standard errors from the Gaussian fits for our fiducial sample. The solid line represents a one-to-one correspondence, while the dashed lines have slopes of 2 and 0.5. Only the simulation and fit errors smaller than $<5\times10^{-3}$ are shown. Right-hand panel: errors estimated from the simulations as a function of redshift. Spectra with $\Delta\left(\Delta\alpha/\alpha\right)_{\rm{fit}}>5\times10^{-3}$ are shown as red crosses ($24\%$ of the total). {\bf There is a clear division between two different set of spectra which correlates with the sky emission lines (see discussion in the main text). }} %The simulation errors are expected to be larger since they are also affected by the continuum subtraction and the systematics as discussed in the text.
\label{fig:sim}
\end{figure*}

\begin{figure*}
\centerline{\includegraphics[angle=0,width=0.5\textwidth]{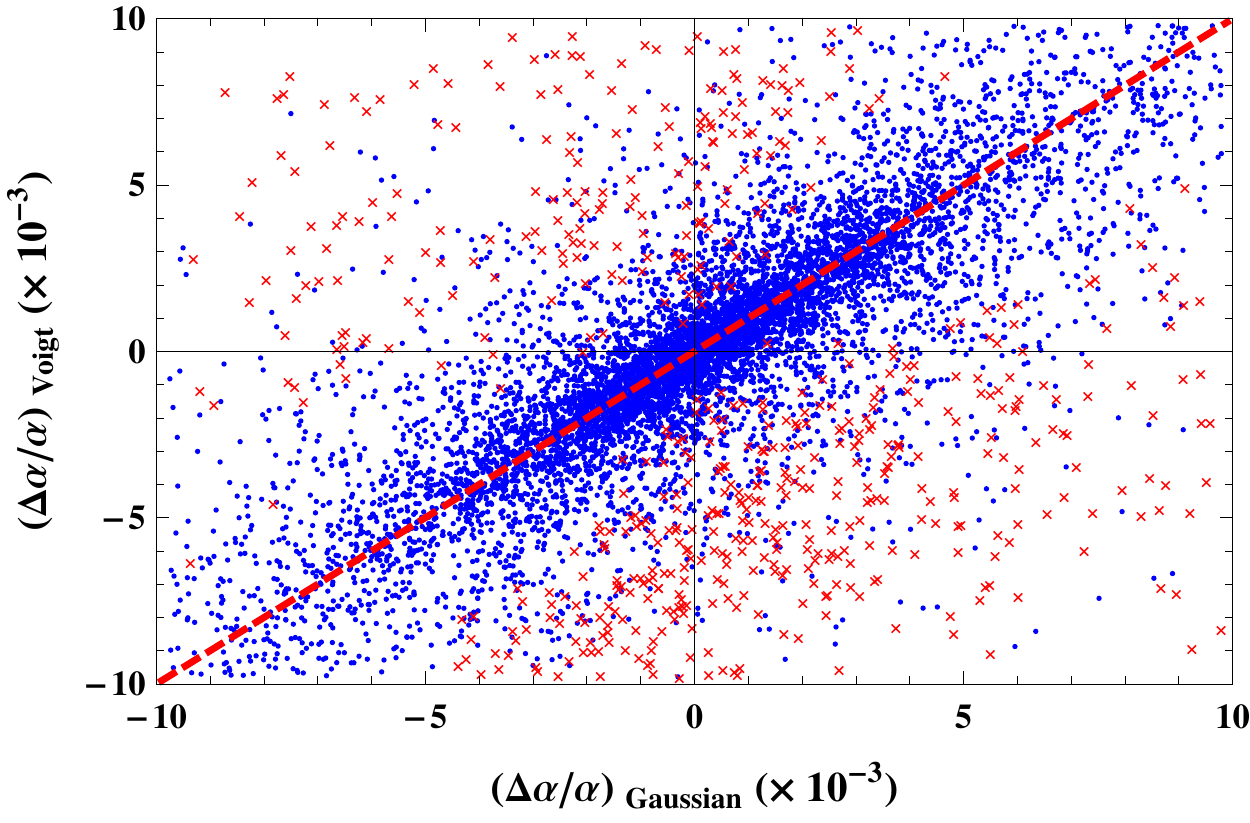}\hfill\includegraphics[angle=0,width=0.5\textwidth]{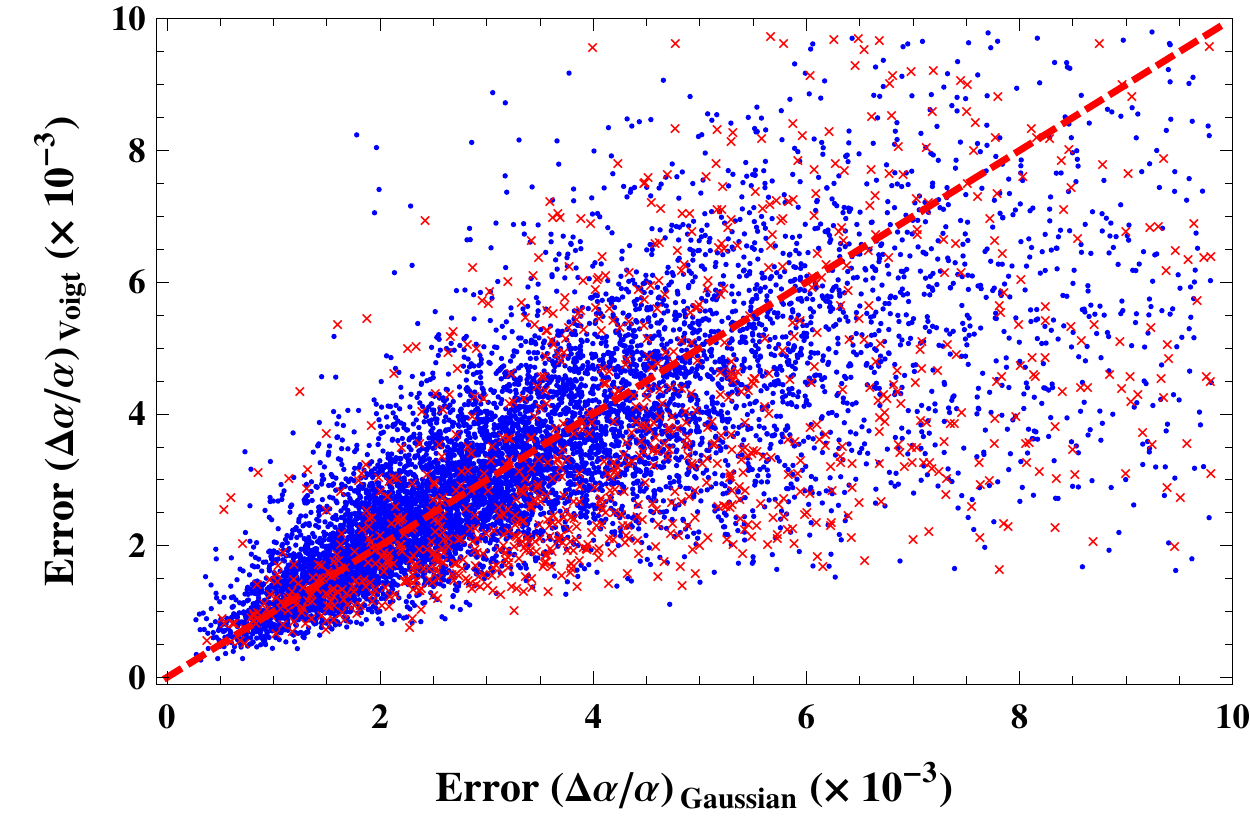}}
\caption{Left-hand panel: measurements of $\Delta\alpha/\alpha$ using Gaussian and Voigt fitting profiles. Non-compatible measurements at 1$\sigma$ are shown as red crosses ($6.5\%$ of the total). Right-hand panel: errors from the Gaussian and Voigt fitting. Non-compatible measurements at 1$\sigma$ are shown as red crosses.}

\label{fig:voigt}
\end{figure*}

In order to test the robustness and accuracy of our methodology, we generate realizations of quasar spectra using as noise a normal distribution centred at the flux value, and taking the error in each pixel as the standard deviation. From our fiducial sample (10\,363 quasars), we simulate 100 realizations for each spectrum ($>$ a million in total). This number of realizations provides reasonable statistics to derive an error from the standard deviation of the measurements on the realizations of each real spectrum, while the computation time remains reasonable ($\sim 2$ d) using a standard-size computer. The estimated error derived from the simulations $\Delta(\Delta\alpha/\alpha)_{\rm{sim}}$ includes
\begin{eqnarray}
\Delta(\Delta\alpha/\alpha)_{\rm{sim}}^2=\Delta(\Delta\alpha/\alpha)_{\rm{fit}}^2+
\Delta(\Delta\alpha/\alpha)_{\rm{continuum}}^2+\Delta(\Delta\alpha/
\alpha)_{\rm{code}}^2
\,,
\end{eqnarray}
where $\Delta(\Delta\alpha/\alpha)_{\rm{fit}}$ is the error derived from the Gaussian fits, which is our error estimate for each real spectrum; $\Delta(\Delta\alpha/\alpha)_{\rm{continuum}}$ is the error from different continuum subtraction due to the Gaussian noise, and $\Delta(\Delta\alpha/\alpha)_{\rm{code}}$ is the systematic error of our code. Then, we expect $\Delta(\Delta\alpha/\alpha)_{\rm{sim}}>\Delta(\Delta\alpha/\alpha)_{\rm{fit}}$ and their difference will be an indication of the continuum and systematic errors.
%Table \ref{tab:simulations} presents a comparison between the value for $\Delta\alpha/\alpha$ obtained from the simulations and from the real spectra of the fiducial sample. 

Fig.\ \ref{fig:sim} (left-hand panel) shows the correlation between the error in $\Delta\alpha/\alpha$ from the Gaussian fits of each real spectrum and the standard deviation for $\Delta\alpha/\alpha$ of its 100 realizations. The standard deviations from the simulations are within a factor of $0.5-2$ of the standard errors from the fits for $97\%$ ($84\%$) of the cases when both quantities are $<5\times10^{-3}$ ($<50\times10^{-3}$). This shows that our code and the continuum subtraction do not introduce noticeable systematic errors compared to the Gaussian fitting. However, there is a set of spectra ($9\%$ of the total) for which the simulations provide much larger errors $\Delta(\Delta\alpha/\alpha)>0.1$. Fig. \ref{fig:sim} (right-hand panel) shows the errors from the simulations as a function of redshift for our fiducial sample. Red crosses stand for spectra whose Gaussian fit error $\Delta(\Delta\alpha/\alpha)_{\rm{fit}}>5\times10^{-3}$ ($24\%$). The errors are distributed in two clouds of points. For the cloud with $\Delta(\Delta\alpha/\alpha)_{\rm sim}\sim 1$, the virtual realizations of each spectrum seem to differ significantly from the real spectrum. Since we use the error in each pixel to build the realizations, the relative error is large for these spectra, which is an indication of a low S/N ratio or large absolute errors in the pixels, for instance in wavelength regions with sky emission lines. In fact, the cloud with bigger errors mimics the sky spectrum. Note also the strong correlation between this cloud of points and the spectra with large Gaussian fitting errors (red crosses). The other set of points with $\Delta(\Delta\alpha/\alpha)_{\rm sim}\sim 10^{-3}$ are close to our error estimation on the measurement of $\Delta\alpha/\alpha$ based on the Gaussian fits.

As a further proof, we also simulate realizations of the 1416 dropped spectra because of sky emission lines (criterion iv, see Section \ref{sec:sample}). We found that more than 80$\%$ of the spectra have $\Delta(\Delta\alpha/\alpha)_{\rm sim}> 0.1$. This confirms that these spectra have very low S/N and/or large pixels error due to the proximity of the lines to strong sky emission lines.

\vspace{-0.5cm}
\subsection{Gaussian versus Voigt fitting profiles}

The results obtained when using Voigt profiles instead of Gaussian ones are compared in Fig.\ \ref{fig:voigt}. The Voigt and Gaussian measurements are $1\sigma$-compatible for the $93.5\%$ of the cases ($98.3\%$ at $2\sigma$). Regarding the errors, there is no clear improvement when using either of both methods. However, Voigt profiles have one more parameter and restrict the number of degrees of freedom. Due to the spectral mid-resolution and the fact that the \oxygen lines are very narrow, there are often only a few pixels to fit, which frequently lead to non-convergent fits. This reduces the quasar sample in $\approx 1000$ quasars. Further discussion about both profiles can be found in Section \ref{sec:systematics}.

%\begin{table}

%\caption{Results for the simulations showing the number of simulations per spectrum, the number of quasar spectra considered and the results for $\Delta\alpha/\alpha$ obtained from the simulations and from the real spectra.}
%\label{tab:simulations}
%\centering
%\begin{tabular}{@{}ccD{,}{\pm}{1.1}D{,}{\pm}{1.1}@{}}%{c | c | c } 
%\midrule[1.8pt] 
%
%$\#$ sim/spec & $\#$ Quasar spectra &   \multicolumn{1}{c}{$\Delta\alpha/\alpha\ (\times 10^{-5})$ sim} & \multicolumn{1}{c}{$\Delta\alpha/\alpha\ (\times 10^{-5})$ real} \\ \midrule[1.8pt]
 	
%100  &  5792 & ??\ ,\ ??
% 	&  ??\ ,\ ??   \\ \midrule[1.8pt]
%
%\end{tabular}

%\end{table}

%%%%%%%%%%%%%%%%%%%%%%%%%%%%%%%%%%%%%%%%%%%%%%%%%%%%%%
%%%%%%%%%%%%%%%%%%%%%%%%%%%%%%%%%%%%%%%%%%%%%%%%%%%%%%

\section{Systematics}
\label{sec:systematics}
%%%%%%%%%%%%%%%%%%%%%%%%%%%%%%%%%%%%%%%%%%%%%%%%%%%%%%
%%%%%%%%%%%%%%%%%%%%%%%%%%%%%%%%%%%%%%%%%%%%%%%%%%%%%%

In this section, we examine the possible unnoticed systematic errors by analysing different quasar samples. Table \ref{tab:samples} summarizes all the samples considered together with their mean redshifts and the measured value for $\Delta\alpha/\alpha$.

We consider the following sources of systematic errors.

\begin{enumerate}

\item{Misidentification of the lines. The expected line widths and amplitudes are useful to avoid misidentification of the \oxygen emission lines. (a) Line widths: since both lines originate on the same upper energy level, their width must coincide. We check that this is the case by considering quasars whose \oxygen line widths are the same within a relative fraction. For more than half of our fiducial sample, the \oxygen line widths differ by less than 10$\%$ (see Table \ref{tab:samples}). \mbox{(b) Amplitude ratio:} atomic physics states that the amplitude ratio between the \oxygen 5008 and \oxygen 4960 lines is 2.98 \citep{Storey} (as quoted in Section \ref{sec:results}, we obtain $2.96\pm0.02_{\rm syst}$). Thus, we consider different samples where this ratio differs by less than a certain amount from $2.98$ (see Table \ref{tab:samples}). All the samples considered in this test yield results for $\Delta\alpha/\alpha$ compatible with zero. \mbox{Fig.\ \ref{fig:widths}} displays the Gaussian widths and fluxes of both \oxygen emission lines for our fiducial sample.}

\item{Windows for the Gaussian fits. We use a wavelength range of 2$\sigma$ around each \oxygen line in order to obtain the final Gaussian fit to the line profiles. We study how our results depend on this choice. By considering a larger wavelength interval, the results are more affected by the H\,$\beta$ contamination and possible asymmetries on the line wings. The differences in the number of spectra for these samples [which are obtained by applying the selection criteria (i)$-$(iv) discussed in Section \ref{sec:sample1}] arise because of the criteria concerning the non-converging fits and the sky emission lines described in Section \ref{sec:sample}.}

\item{H\,$\beta$ contamination. We analyse samples where the ratio between S/N$_{\rm{H}\,\beta}$ and S/N$_{\rm{\oxygen 4960}}$ is constrained. Despite the fact that the value for $\Delta\alpha/\alpha$ decreases as we place more stringent constraints on H\,$\beta$, it is always consistent with no variation in $\alpha$ within the errors. This analysis demonstrates that the strength and/or width of the H\,$\beta$ line do not affect substantially the result for $\Delta\alpha/\alpha$ when a weighted mean is adopted.}

\item{Continuum subtraction. We use a seventh-order polynomial to subtract the continuum spectrum. We examine if the polynomial order has important effects on our measurements. Our values for $\Delta\alpha/\alpha$ and their errors are only slightly affected by the chosen polynomial order.}

\item{Goodness of Gaussian fits. We quantify the quality of the Gaussian fits by the $R^2$ coefficient. All the considered samples show values for $\Delta\alpha/\alpha$ consistent with no variation in $\alpha$.}

\item{Broad lines. We also study samples where the width of both lines is less than a certain value (in km\,s$^{-1}$). These samples are consistent with no variation of $\alpha$. Samples built from narrow lines $<300$ km\,s$^{-1}$} may be more affected by misidentification of \oxygen lines as sky lines.

\begin{table}

\caption{Results for $\Delta\alpha/\alpha$ considering several samples with different constraints. The number of quasar spectra, the mean and standard deviation of the redshift and the value for $\Delta\alpha/\alpha$ are shown.}
\label{tab:samples}
\centering

\begin{tabular}{@{}c@{\hskip 0.1cm}c@{\hskip 0.cm}D{,}{\pm}{1.1}@{\hskip -0.3cm}D{,}{\pm}{1.1}@{}}%{c | c | c } 
\midrule[1.8pt] 
$\sigma_{4960/5008}-1$ & No.\ of quasar spectra &  \multicolumn{1}{c}{Redshift}&  \multicolumn{1}{c}{$\Delta\alpha/\alpha\ (\times 10^{-5})$} \\ \midrule[0.7pt]

$<50\%$  &  $10\,028$ & 0.56\ ,\ 0.21
 	&  1.6\ ,\ 2.3  \\
 	
$<25\%$  &  $8877$ & 0.56\ ,\ 0.21
 	&  1.9\ ,\ 2.3  \\
 	
$<10\%$  &  $5846$ & 0.56\ ,\ 0.21
 	&  1.7\ ,\ 2.5  \\
 	
$<5\%$  &  $3458$ & 0.54\ ,\ 0.22
 	&  -0.9\ ,\ 3.0  \\ \midrule[1.8pt]

$\left[F_{\lambda}\times\sigma\right]_{5008/4960}$ & No.\ of quasar spectra &  \multicolumn{1}{c}{Redshift}&  \multicolumn{1}{c}{$\Delta\alpha/\alpha\ (\times 10^{-5})$} \\ \midrule[0.7pt]

$2.98\pm0.50$  &  $8327$ & 0.56\ ,\ 0.21 
 	&  1.8\ ,\ 2.4  \\
 	
$2.98\pm0.25$  &  $5761$ & 0.55\ ,\ 0.21
 	&  -0.4\ ,\ 2.6  \\
 	
$2.98\pm0.10$  &  $2658$ & 0.54\ ,\ 0.21
 	&  0.0\ ,\ 3.4  \\
 	
$2.98\pm0.05$  &  $1411$ & 0.52\ ,\ 0.22
 	&  5.2\ ,\ 4.6 \\ \midrule[1.8pt]
 	 	
 	Fit width & No.\ of quasar spectra &  \multicolumn{1}{c}{Redshift}&  \multicolumn{1}{c}{$\Delta\alpha/\alpha\ (\times 10^{-5})$} \\ \midrule[0.7pt]

$2\sigma$  &  $10\,363$ & 0.56\ ,\ 0.21
 	&  1.4\ ,\ 2.3  \\ 
 	
$3\sigma$  &  $10252$ & 0.59\ ,\ 0.20
 	&  5.5\ ,\ 2.5  \\
 	
$4\sigma$  &  $9978$ & 0.59\ ,\ 0.20
 	&  7.1\ ,\ 2.7  \\
 	
$5\sigma$  &  $9726$ & 0.59\ ,\ 0.20
 	&  5.3\ ,\ 2.6   \\ \midrule[1.8pt]
 	
S/N$_{\rm{H\,\beta}/\rm{[O{\sc iii}] 4960}}$ & No.\ of quasar spectra &  \multicolumn{1}{c}{Redshift}&  \multicolumn{1}{c}{$\Delta\alpha/\alpha\ (\times 10^{-5})$} \\ \midrule[0.7pt]

$<5$  &  $10\,338$ & 0.57\ ,\ 0.21
 	&  1.4\ ,\ 2.3  \\ 
 	
$<2$  &  $9831$ & 0.57\ ,\ 0.21
 	&  0.6\ ,\ 2.3  \\
 	
$<1$  &  $8162$ & 0.57\ ,\ 0.21
 	&  0.1\ ,\ 2.5  \\
 	
$<0.5$  &  $5831$ & 0.58\ ,\ 0.21
 	&  -0.7\ ,\ 2.8   \\ \midrule[1.8pt]

Pol. order (cont.) & No.\ of quasar spectra &  \multicolumn{1}{c}{Redshift}&  \multicolumn{1}{c}{$\Delta\alpha/\alpha\ (\times 10^{-5})$} \\ \midrule[0.7pt]

$3$  &  $10\,528$ & 0.57\ ,\ 0.21
 	&  1.0\ ,\ 2.3 \\ 
 	
$5$  &  $10\,550$ & 0.57\ ,\ 0.21
 	&  1.3\ ,\ 2.3 \\
 	
$7$  &  $10\,363$ & 0.56\ ,\ 0.21
 	&  1.4\ ,\ 2.3 \\
 	
$9$  &  $10\,471$ & 0.56\ ,\ 0.21
 	&  -1.1\ ,\ 2.3 \\ \midrule[1.8pt]

$R^2$ (both fits) & No.\ of quasar spectra &  \multicolumn{1}{c}{Redshift}&  \multicolumn{1}{c}{$\Delta\alpha/\alpha\ (\times 10^{-5})$} \\ \midrule[0.7pt]

$>0.9$  &  $9254$ & 0.56\ ,\ 0.21
 	&  1.5\ ,\ 2.4  \\
 	
$>0.97$  &  $6045$ & 0.56\ ,\ 0.21
 	&  2.8\ ,\ 2.7  \\
 	
$>0.99$  &  $2301$ & 0.54\ ,\ 0.21
 	&  2.0\ ,\ 3.5  \\
 	
$>0.995$  &  $845$ & 0.51\ ,\ 0.22
 	&  -0.4\ ,\ 4.8  \\ \midrule[1.8pt]

\oxygen$_{5008}$ (km\,s$^{-1}$)& No.\ of quasar spectra &  \multicolumn{1}{c}{Redshift} &  \multicolumn{1}{c}{$\Delta\alpha/\alpha\ (\times 10^{-5})$} \\ \midrule[0.7pt]

$<1000$  &  $10\,353$ & 0.56\ ,\ 0.21
 	&  1.4\ ,\ 2.3  \\
 	
$<500$  &  $8990$ & 0.56\ ,\ 0.21
 	&  0.2\ ,\ 2.4  \\
 	
$<300$  &  $2798$ & 0.52\ ,\ 0.22
 	&  -6.8\ ,\ 3.9  \\
 	
$<200$  &  $150$ & 0.52\ ,\ 0.24
 	&  21\ ,\ 18  \\ \midrule[1.8pt]
 	
 	Method & No.\ of quasar spectra &  \multicolumn{1}{c}{Redshift}&  \multicolumn{1}{c}{$\Delta\alpha/\alpha\ (\times 10^{-5})$} \\ \midrule[0.7pt]

 Gaussian (weighted) &  $4537$ & 0.58\ ,\ 0.20
 	&  -0.4\ ,\ 2.8  \\
 	 	
 Gaussian &  $4537$ & 0.58\ ,\ 0.20
 	&  1.2\ ,\ 4.5 \\
 	
 Integration &  $4537$ & 0.58\ ,\ 0.20
 	&  3.6\ ,\ 4.8 \\

 Modified Bahcall &  $4537$ & 0.58\ ,\ 0.20
 	&  0.8\ ,\ 4.4 \\
 	
  Median &  $4537$ & 0.58\ ,\ 0.20
 	&  1.8\ ,\ 1.4  \\

 \midrule[1.8pt]

  	Gauss versus Voigt & No.\ of quasar spectra &  \multicolumn{1}{c}{Redshift}&  \multicolumn{1}{c}{$\Delta\alpha/\alpha\ (\times 10^{-5})$} \\ \midrule[0.7pt]

 Gaussian profiles &  $8485$ & 0.55\ ,\ 0.19
 	&  0.4\ ,\ 2.5  \\
 	 	
 Voigt profiles &  $8485$ & 0.55\ ,\ 0.19
 	&  -1.1\ ,\ 2.8 \\
 	
 Mixed profiles &  $8485$ & 0.55\ ,\ 0.19
 	&  1.3\ ,\ 2.4 \\

 \midrule[1.8pt]

 	%Spectrograph channel & $\#$ Quasar spectra &  \multicolumn{1}{c}{redshift}&  \multicolumn{1}{c}{\hspace{0.25cm}$\Delta\alpha/\alpha\ (\times 10^{-5})$} \\ \midrule[1.8pt]

%Blue ($z<0.2$)  &  $149$ & 0.13\ ,\ 0.05  
 %	&  2.2\ ,\ 9.6  \\ 
 	
%Red ($z>0.2$)  &  $10,214$ & 0.59\ ,\ 0.19
% 	&  1.3\ ,\ 2.4  \\ \midrule[1.8pt]

%Degrees of freedom & $\#$ Quasar spectra &  \multicolumn{1}{c}{redshift}&  \multicolumn{1}{c}{\hspace{0.25cm}$\Delta\alpha/\alpha\ (\times 10^{-5})$} \\ \midrule[1.8pt]

%$1\leq dof\leq3$  &  $260$ & 0.48\ ,\ 0.23
 %	&  23\ ,\ 13  \\
 	
%$4\leq dof \leq 6$  &  $3,459$ & 0.55\ ,\ 0.21
 %	&  1.0\ ,\ 3.6  \\
 	
%$7\leq dof \leq 9$  &  $3,743$ & 0.61\ ,\ 0.19
 %	&  11\ ,\ 4  \\
 	
%$dof \geq 10$  &  $1,862$ & 	0.64\ ,\ 0.18
 %	&  24\ ,\ 7  \\ \midrule[1.8pt]

\end{tabular}

\end{table}

\begin{figure*}
\centerline{\includegraphics[angle=0,width=0.49\textwidth]{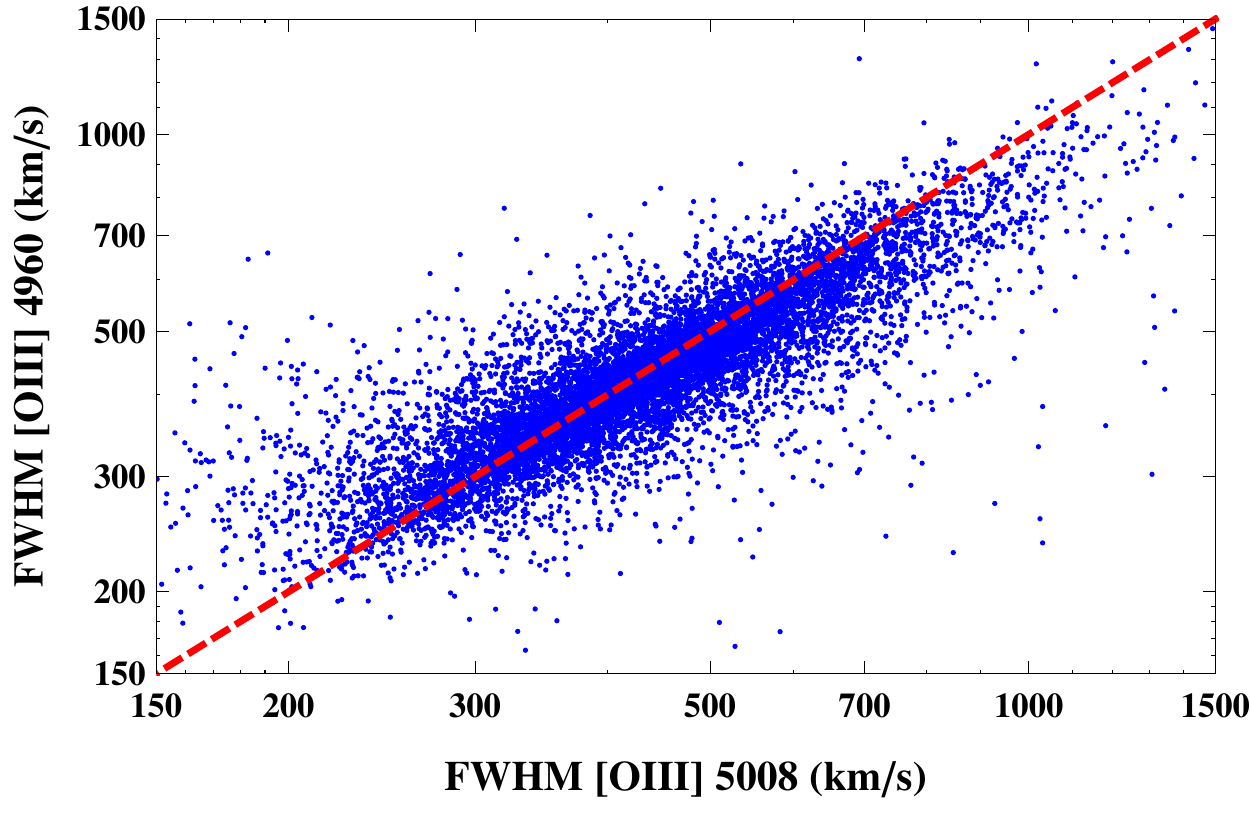}\hfill\includegraphics[angle=0,width=0.49\textwidth]{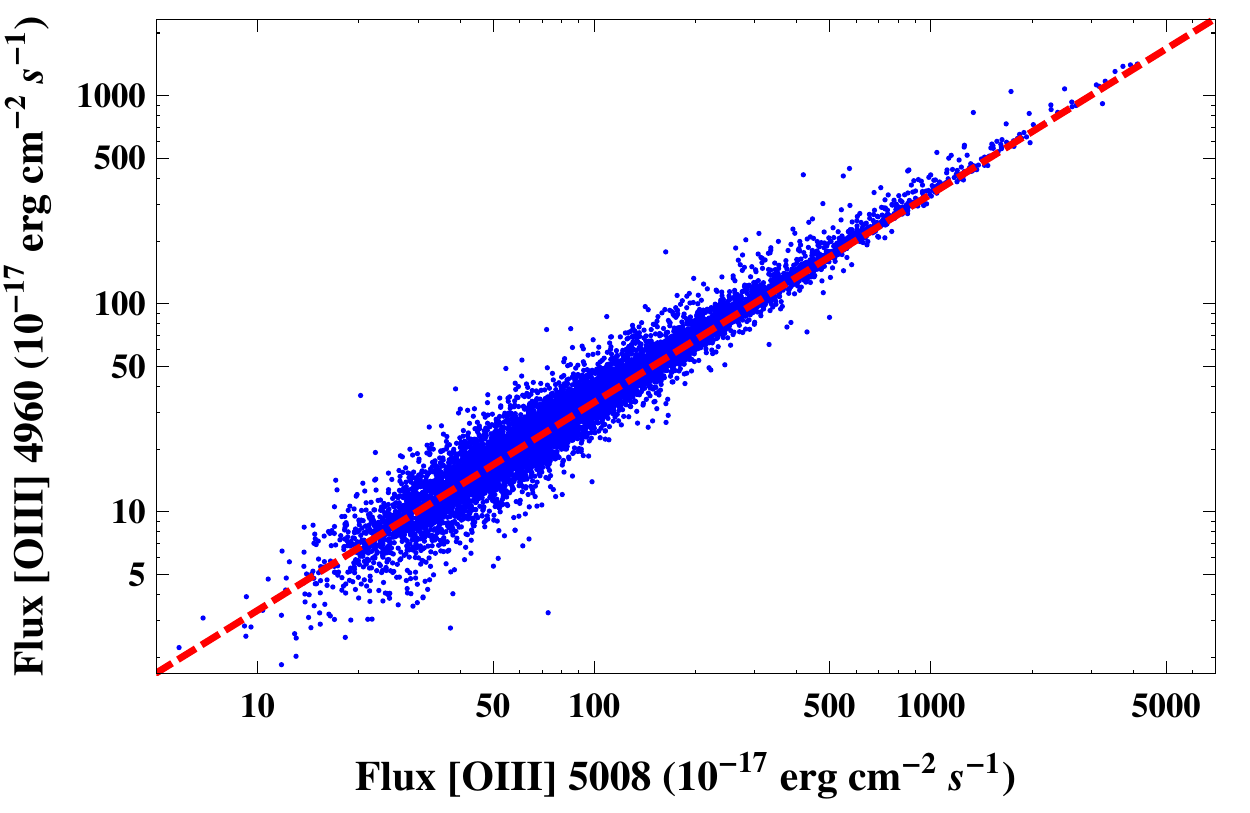}}
\caption{Left-hand panel: Gaussian widths (in km\,s$^{-1}$) for both \oxygen lines. Both lines originate on the same upper energy level, then their widths must coincide (red dashed line). Right-hand panel: fluxes for both \oxygen lines. The theoretical flux ratio is $2.98$ (red dashed line). The entire fiducial quasar sample is shown.}
\label{fig:widths}
\end{figure*}

\item{Different methods for measuring the \oxygen line position. We compare the results obtained by the methods to measure the position of the \oxygen lines described in Section \ref{sec:methodology24}. Since not all the methods provide an error for the measurement, we cannot calculate a weighted mean, and it is necessary to select a more restricted sample. Then, we consider a sample where the difference between the widths of the lines is less than $25\%$, the amplitude ratio is constrained to differ from the theoretical value 2.98 \citep{Storey} by less than 0.5, and the S/N$_{\rm{H}\,\beta}$ is smaller than half the S/N$_{\rm{\oxygen 4960}}$.}

\item{Gaussian versus Voigt profiles. We compare the results for 8485 quasars from our fiducial sample after dropping 1878 spectra with non-converging Voigt fits (this reduction increases the statistical error). We also compute a `mixed' value for $\Delta\alpha/\alpha$ where for each spectrum we use the value for the variation of the fine-structure constant with smaller error, either  $(\Delta\alpha/\alpha)_{\rm{Gauss}}$ or $(\Delta\alpha/\alpha)_{\rm{Voigt}}$.}

\end{enumerate}

\begin{figure}
\centerline{\includegraphics[angle=0,width=0.48\textwidth]{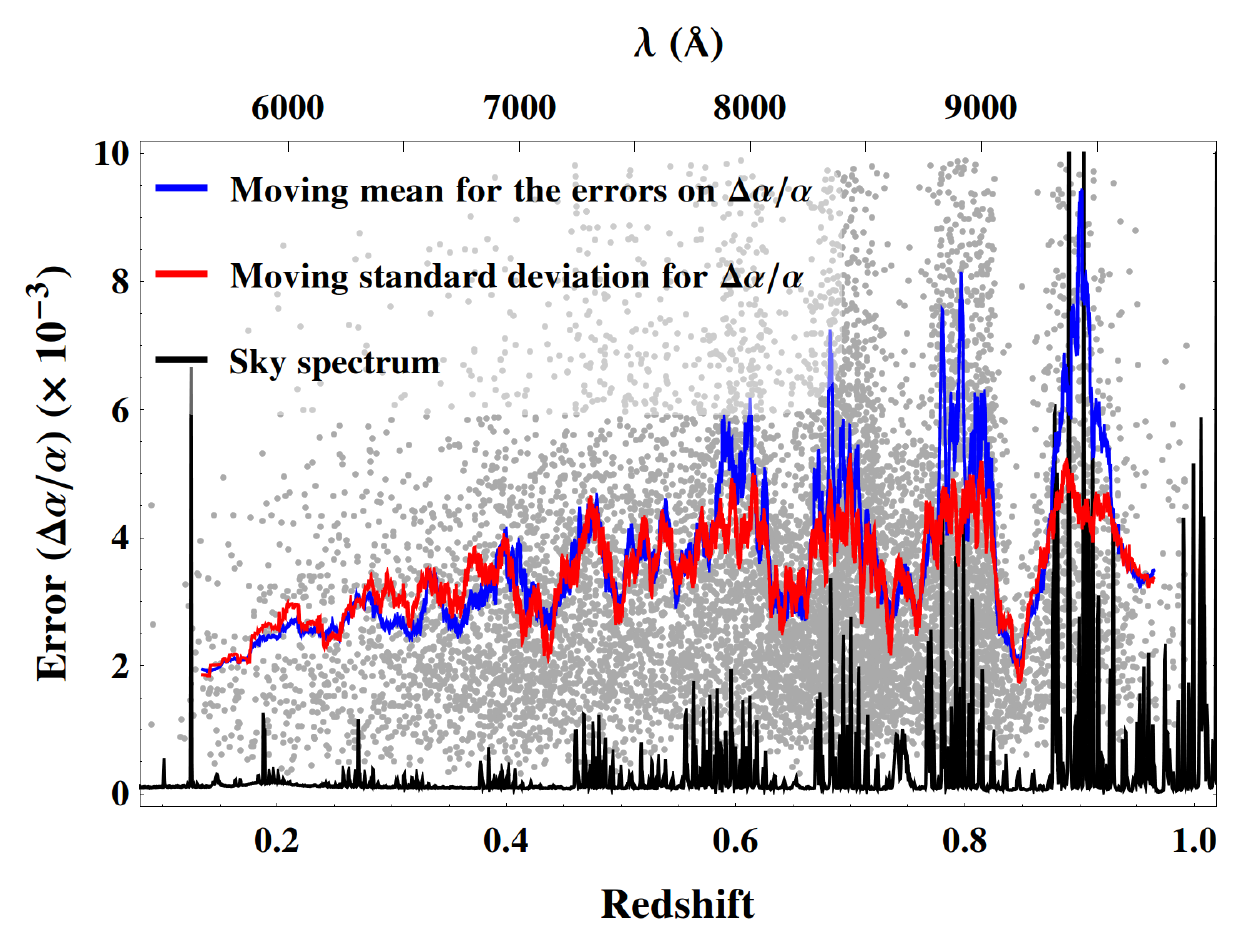}}
\caption{Errors for $\Delta\alpha/\alpha$ derived from the Gaussian fits (grey points) for our fiducial sample, moving mean of the these errors (blue line) using overlapping bins (100 spectra per bin, $\Delta\, z\approx0.025$), moving standard deviation of $\Delta\alpha/\alpha$ measurements using the same bins (red line) and a typical sky spectrum.}
%The correlation between the sky; and the standard deviation and the fit errors for the $\Delta\alpha/\alpha$ measurement is clear. Notice that this correlation is expected for the Gaussian fit errors (black line) since they are affected by the errors in each pixel which are affected in turn by the presence of sky lines. However, it is quite surprising that we can build the sky from the standard deviation (red line), which does not content any information about it. This means that the precision in our determination of the line positions (and consequently in our result for $\Delta\alpha/\alpha$) is limited by the SDSS sky subtraction algorithm.
\label{fig:sky}
\end{figure}

We have also analysed the standard deviation and errors of the results for $\Delta\alpha/\alpha$ as a function of redshift (Fig.\ \ref{fig:sky}). Even though we have imposed a constraint on our initial sample based on the sky emission lines, the standard deviation and errors still correlate with the sky. In particular, for the correlation with the moving standard deviation, this means that the precision in our measurement of $\Delta\alpha/\alpha$ along the whole redshift interval is limited by the sky subtraction algorithm.
%The redshift interval for reach bin depends on the density of quasar as a function of redshift but it is approximately $\Delta z\approx 0.025$. These overlapping bins act as an effective low-band pass filter, showing the dispersion of the errors and standard deviation of the measurements over scales $>$0.025 in redshift. They also provide a clean

\vspace{-0.5cm}

%%%%%%%%%%%%%%%%%%%%%%%%%%%%%%%%%%%%%%%%%%%%%%%%%%%%%%
%%%%%%%%%%%%%%%%%%%%%%%%%%%%%%%%%%%%%%%%%%%%%%%%%%%%%%

\section{Results}
\label{sec:results}
%%%%%%%%%%%%%%%%%%%%%%%%%%%%%%%%%%%%%%%%%%%%%%%%%%%%%%
%%%%%%%%%%%%%%%%%%%%%%%%%%%%%%%%%%%%%%%%%%%%%%%%%%%%%%

\subsection{\oxygen lines}

\begin{figure*}
\centerline{\includegraphics[angle=0,width=0.495\textwidth]{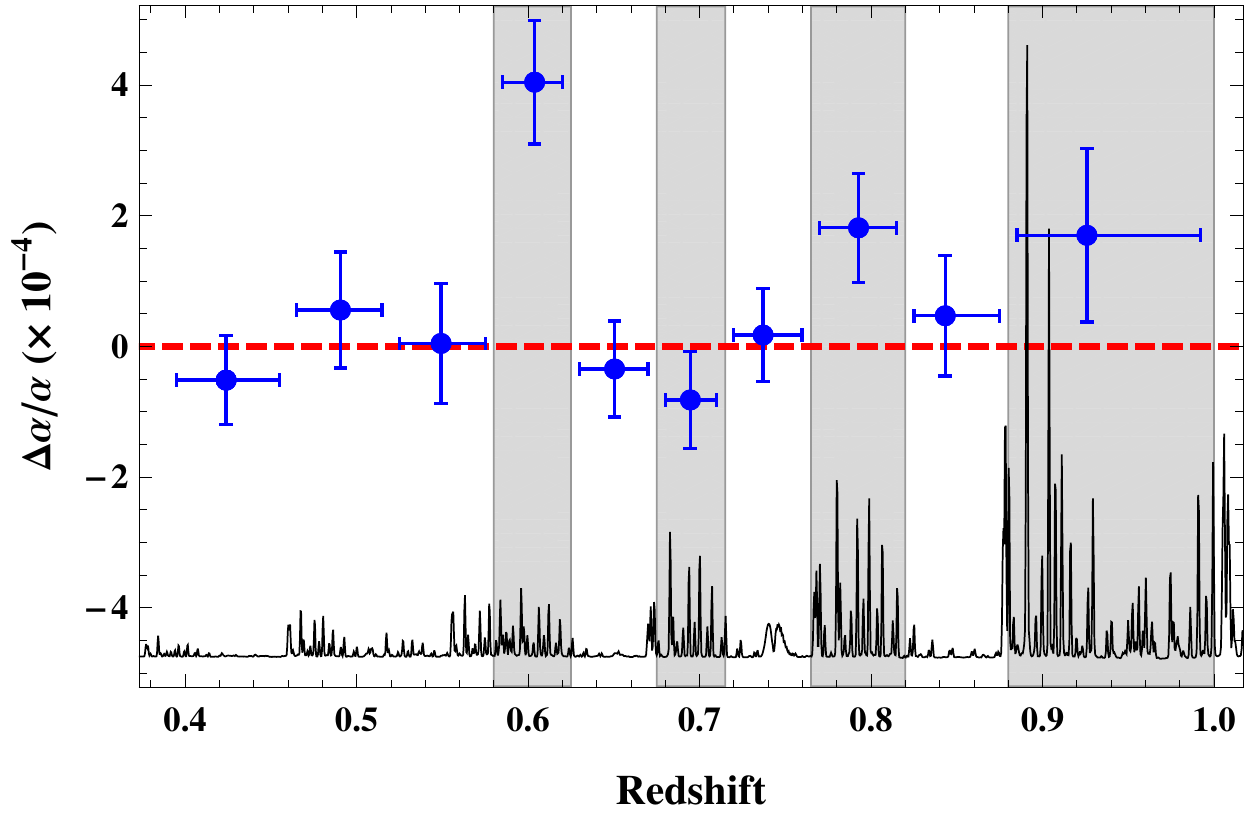}\hfill\includegraphics[angle=0,width=0.495\textwidth]{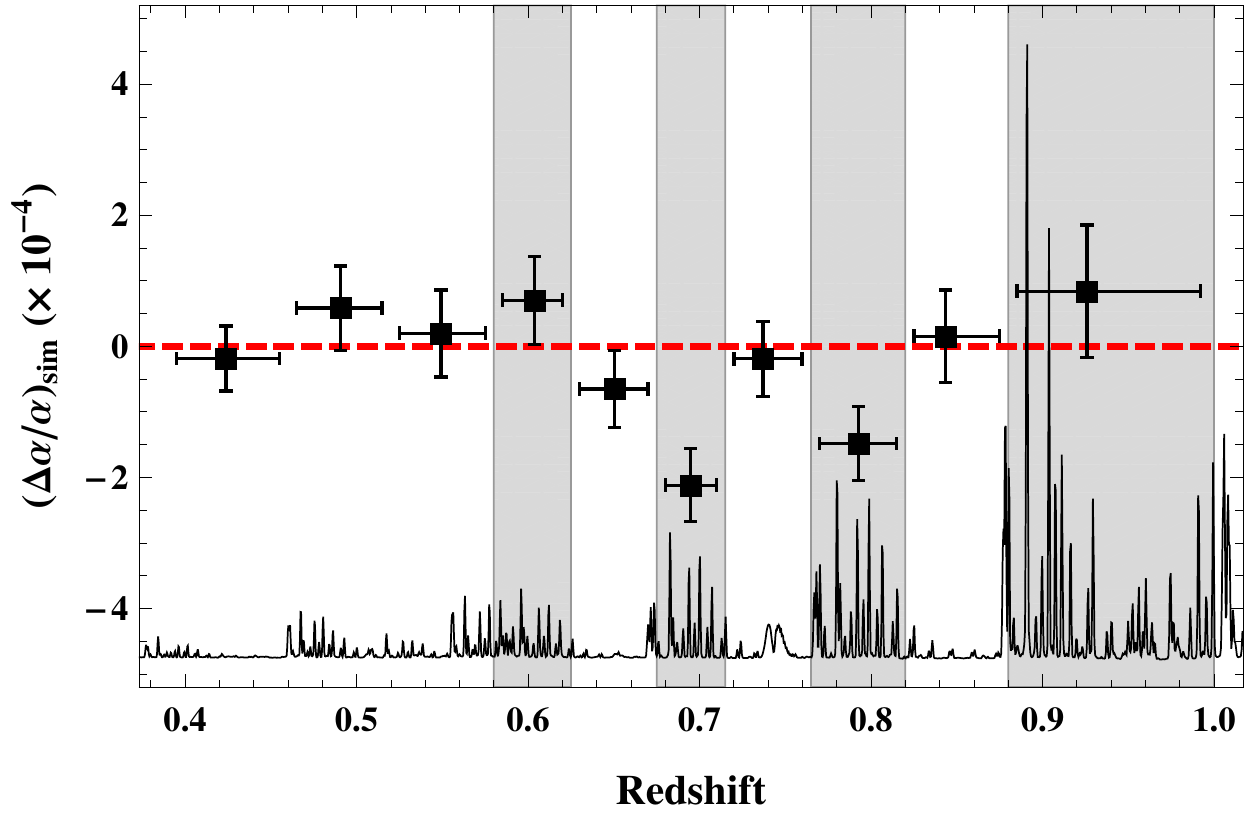}}
\caption{Left-hand panel: $\Delta\alpha/\alpha$ versus redshift (real data). Details about each bin are listed in Table \ref{tab:bins}. Right-hand panel: $(\Delta\alpha/\alpha)_{\rm{sim}}$ versus redshift (simulations). A typical sky spectrum and shadowed regions where the sky contamination is expected to be large, are shown as reference.}
\label{fig:alpha}
\end{figure*}

We used a total of 10\,363 quasar spectra, drawn from the SDSS-III/BOSS DR12Q catalogue, after applying the selection criteria (i)$-$(iv) (see Section \ref{sec:sample}), to measure the possible variation of the fine-structure constant. The following measurement is obtained:
\begin{eqnarray}
\frac{\Delta\alpha}{\alpha}=\left(1.4 \pm 2.3 \right) \times 10^{-5} \nonumber\,.
\end{eqnarray}
This value is consistent with the previous results reported in different investigations based on the same method: \citet{Bahcall}, \citet{Gutierrez}, and \citet{Rahmani}. The redshift dependence of the measurements is shown in Fig.\ \ref{fig:alpha} (left-hand panel), where several bins have been made taking into account the redshift intervals affected by the sky (shaded zones). In the right-hand panel, we show the results obtained from the simulations described in Section \ref{sec:methodology}, using the same redshifts intervals for the bins. The main differences between the real results and the simulations are in the regions where there are strong sky lines (shaded regions), while being in agreement in the remaining zones. Detailed information about each bin for the real data can be found in \mbox{Table \ref{tab:bins}.}

Our results are little affected by the specific constraints imposed in our sample as discussed in Section \ref{sec:systematics}. For instance, we vary the width for the Gaussian fits, the contamination of H\,$\beta$, the polynomial order used to fit the continuum spectrum, the quality of the Gaussian fits and test different methods to measure $\Delta\alpha/\alpha$. The most important effect found is that by considering broader widths for the Gaussian fits, the results are more affected by the contamination from H\,$\beta$ and possible asymmetries in the line wings. We have also checked for possible misidentifications of the \oxygen emission lines using their expected widths and amplitude ratio.

\begin{figure}
\centerline{\includegraphics[angle=0,width=0.49\textwidth]{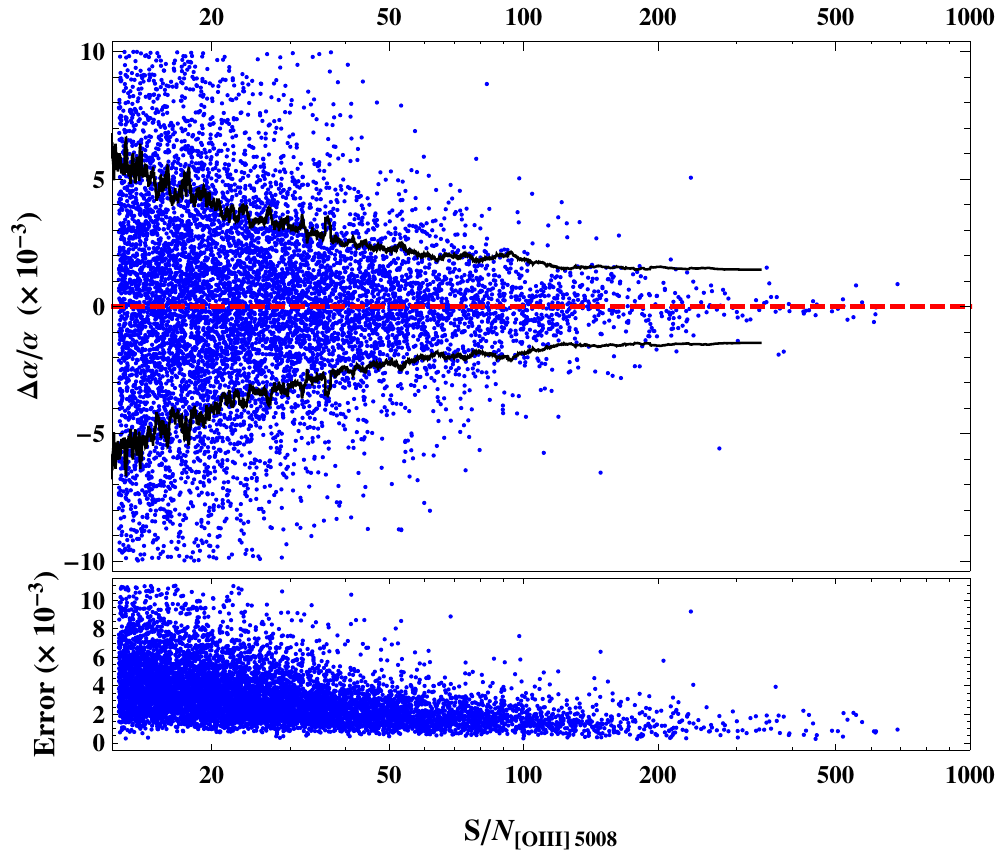}}
\caption{$\Delta\alpha/\alpha$ versus S/N$_{\rm{\oxygen 5008}}$ (top panel) with the moving standard deviation (black lines) using overlapping bins (100 spectra per bin) and the error on $\Delta\alpha/\alpha$ (bottom panel) in linear-log scale for our fiducial sample. The deviation of $\Delta\alpha/\alpha$ from zero and its error steadily decreases as the S/N increases.}
\label{fig:SN}
\end{figure}

Table \ref{tab:results} contains the results for $\Delta\alpha/\alpha$ when the lower bound on the S/N$_{\rm{\oxygen 5008}}$ is increased. All the results remain consistent with no variation of the fine-structure constant. In Fig.\ \ref{fig:SN}, the measured $\Delta\alpha/\alpha$ for our fiducial sample as a function of the S/N$_{\rm{\oxygen 5008}}$ are plotted together with their errors.

\begin{table}

\caption{Detailed information about the bins in Fig.\ \ref{fig:alpha}.}
\label{tab:bins}
\centering
\begin{tabular}{@{}c@{\hskip 0.2cm}c@{\hskip 0.2cm}D{,}{\pm}{4.4}D{,}{\pm}{4.4}@{}}%{c | c | c } 
\midrule[1.8pt] 
Redshift interval & No. of quasar spectra &  \multicolumn{1}{c}{Redshift\ \ \ \ } &  \multicolumn{1}{c}{$\Delta\alpha/\alpha\ (\times 10^{-5})$} \\ \midrule[0.7pt] 
 
$0.390-0.460$ &  $817$ & 0.42\ ,\ 0.02
 	&  -5.2\ ,\ 6.8  \\

$0.460-0.520$  &  $723$  & 0.49\ ,\ 0.02
 	&  5.5\ ,\ 8.9  \\ 
 	
$0.520-0.580$  &  $757$  & 0.55\ ,\ 0.02
 	&  0.4\ ,\ 9.2  \\ 
 	
$0.580-0.625$  &  $843$  & 0.60\ ,\ 0.01
 	&  40.4\ ,\ 9.4  \\ 
    
$0.625-0.675$  &  $988$  & 0.65\ ,\ 0.01
 	&  -3.5\ ,\ 7.4 \\
 	
$0.675-0.715$  &  $1299$  & 0.69\ ,\ 0.01
 	&  -8.2\ ,\ 7.4 \\
 	
$0.715-0.765$  &  $1117$  & 0.74\ ,\ 0.01
 	&  1.7\ ,\ 7.1 \\
 	
$0.765-0.820$  &  $1444$  & 0.79\ ,\ 0.02
 	&  18.1\ ,\ 8.3 \\
 	
$0.820-0.880$  &  $644$  & 0.84\ ,\ 0.02
 	&  4.7\ ,\ 9.2  \\
 	
$0.880-1.000$  &  $580$  & 0.93\ ,\ 0.03
 	&  17.0\ ,\ 13.3  \\
\midrule[1.8pt] 

\end{tabular}

\end{table}

\begin{table}

\caption{Results for several samples with different constraints on the S/N$_{\rm{\oxygen 5008}}$. For each sample, the number of quasar spectra, the mean redshift, together with its standard deviation and the value for $\Delta\alpha/\alpha$ are shown.}
\label{tab:results}
\centering
\begin{tabular}{@{}ccD{,}{\pm}{4.4}D{,}{\pm}{4.4}@{}}%{c | c | c } 
\midrule[1.8pt] 
S/N$_{\rm{\oxygen 5008}}$ & No.\ of quasar spectra &  \multicolumn{1}{c}{Redshift} &  \multicolumn{1}{c}{$\Delta\alpha/\alpha\ (\times 10^{-5})$} \\ \midrule[0.7pt] 
 
$>10$  &  $10\,363$ & 0.56\ ,\ 0.21
 	&  1.4\ ,\ 2.3  \\

$>20$  &  $5270$  & 0.53\ ,\ 0.21
 	&  -0.5\ ,\ 2.5   \\ 
 	
$>50$  &  $1498$  & 0.47\ ,\ 0.20
 	&  -3.4\ ,\ 3.1  \\ 
    
$>100$  &  $451$  & 0.41\ ,\ 0.19
 	&  -2.0\ ,\ 3.6  \\
 	
$>500$  &  $12$  & 0.24\ ,\ 0.19
 	&  6\ ,\ 12  \\
\midrule[1.8pt] 

\end{tabular}

\end{table}

The distribution of BOSS quasars in the sky (see Fig.\ \ref{fig:QSO}, left-hand panel) suggests to divide the sample into two, one for the North galactic cap and one for the South galactic cap. Table \ref{tab:hemispheres} describes the results for each galactic hemisphere, and no statistical meaningful difference is found. In order to look for a spatial variation, we also carried out a more precise analysis by fitting a dipole. First, we fixed the direction on the sky of the dipole and performed a linear fit ($\Delta\alpha/\alpha(\theta)=a\cos{\theta}+b$) of the measurements of the variation of the fine-structure constant as a function of $\cos{\theta}$, where $\theta$ is the angle between the dipole axis and a quasar in the sky. Different fits were done for the dipole axis lying in a grid in RA and Dec\,. ($\sim 1\degree\times1\degree$). However, there is not statistical significance to determine the dipole axis with a meaningful error, i.e.\ smaller than the whole sky. There has been a claim on a significant deviation of $\alpha$ from being a constant as a function of space \citep{King}, with a dipole amplitude $0.97^{+0.22}_{-0.20}\times 10^{-5}$ in the direction RA$=17.3\pm1.0$\,h and Dec.$=-61\degree\pm10\degree$. Fixing the dipole in that direction, we get $\left(-4.3\pm3.4\right)\times10^{-5}$ for the dipole amplitude and $\left(0.8\pm2.3\right)\times10^{-5}$ for the monopole term, which are not precise enough to compare with that work.

\begin{table}

\caption{Results for the North and South galactic hemispheres.}
\label{tab:hemispheres}
\centering
\begin{tabular}{@{}c@{\hspace{0.1cm}}c@{\hspace{0.1cm}}D{,}{\pm}{2.2}D{,}{\pm}{2.2}@{}}%{c | c | c } 
\midrule[1.8pt] 
Galactic hemisphere & No. of quasar spectra &  \multicolumn{1}{c}{Redshift\ \ \ \ }&\multicolumn{1}{c}{$\Delta\alpha/\alpha\ (\times 10^{-5})$} \\ \midrule[0.7pt]

North  &  $8069$ & 0.56\ ,\ 0.21 
 	&  2.6\ ,\ 2.6  \\ 
 	
South  &  $2294$ & 0.59\ ,\ 0.20
 	&  -3.1\ ,\ 4.9   \\ \midrule[1.8pt]
%
%
%Black hole mass & $\#$ Quasar spectra &  \multicolumn{1}{c}{redshift}&  \multicolumn{1}{c}{\hspace{0.25cm}$\Delta\alpha/\alpha\ (\times 10^{-5})$} \\ \midrule[1.8pt]

%$<10^{8.7}\,M_{\odot}$  &  $44$ & 0.39\ ,\ 0.20 
% 	&  -7\ ,\ 20  \\ 
 	
%$>10^{8.7}\,M_{\odot}$  &  $44$ & 0.66\ ,\ 0.25
%	&  30\ ,\ 31   \\ \midrule[1.8pt]
%
\end{tabular}

\end{table}

We are inclined to parametrize the possible time variation of $\alpha$ with redshift $z$. This is justified since any possible variation on $\alpha$ must be dominated by the local geometry of space-time (at least if we consider the dynamics of the Universe as the main reason for such variation). Therefore, one is led to consider the possible variation of $\alpha$ as a function of redshift ($z=1/a(t)-1$) or the Ricci scalar ($R(t)=6H(t)^2[1-q(t)]$), where $a(t)$ is the scale factor, $H(t)$ the {\it Hubble} parameter and $q(t)$ is the deceleration parameter. Since the Ricci scalar is not known for each quasar, it is straightforward to consider a possible variation with redshift. In contrast, for a time parametrized model of the variation of $\alpha$ the analysis depends on the particular cosmology considered. Since there is no significant clear dependence, we use a linear model in redshift. Then, for
\begin{equation}
\Delta\alpha/\alpha=a\,z+b\,,
\end{equation}
we obtain
\begin{equation}
a=\left(0.7\pm2.1\right)\times 10^{-4}\,;\hspace{0.5cm}b=\left(0.7\pm1.4\right)\times10^{-4}\,;
\end{equation}
which do not show any dependence of $\Delta\alpha/\alpha$ with redshift.

From this sample, we also obtain a value for the line ratio $\left[F_{\lambda}\times\sigma\right]_{5008}
/\left[F_{\lambda}\times\sigma\right]_{4960}=2.96\pm0.02_{\rm syst}$, where $F_{\lambda}$ is the maximum flux density of the line, and $\sigma$ is the Gaussian width. The value reported is a weighted mean where the S/N$_{\rm{\oxygen 5008}}$ is used as weights. The quoted systematic error is computed from the analysis of samples with different polynomial orders for the continuum fit and different widths for the line fitting (see Table \ref{tab:amplitude}), since this quantity is more affected by these two parameters. The value we obtain is in agreement with %previous estimates using the same method, \citet{Bahcall} ($2.99\pm0.02$) and \citet{Gutierrez} ($3.012\pm0.01$); \citet{Gutierrez} ($3.012\pm0.01$); 
the best current theoretical value, i.e., $2.98$ \citep{Storey}.% and from active galactic nuclei measurements \citet{AGN} ($2.993\pm0.014$).

\begin{table}

\caption{Results for the line ratio when polynomials of different orders are used to subtract the continuum and different range for the Gaussian fits are used. For each sample, the number of quasar spectra, the mean redshift together with its standard deviation and the value for $\left[F_{\lambda}\times\sigma\right]_{5008}
/\left[F_{\lambda}\times\sigma\right]_{4960}$ are shown.}
\label{tab:amplitude}
\centering
\begin{tabular}{@{}c@{\hspace{0.15cm}}c@{\hspace{0.1cm}}D{,}{\pm}{2.2}@{\hspace{0.05cm}}c@{}}%{c | c | c } 
\midrule[1.8pt] 
Polynomial order & No.\  of quasar spectra &  \multicolumn{1}{c}{Redshift\ \ } & $\left[F_{\lambda}\times\sigma\right]_{5008
/4960}$ \\ \midrule[0.7pt] 
 
3  &  $10\,528$ & 0.57\ ,\ 0.21 
 	& 2.96 \\

5  &  $10\,550$  & 0.57\ ,\ 0.21
 	&  2.94 \\ 
 	
7  &  $10\,363$  & 0.56\ ,\ 0.21 
 	&  2.96 \\ 
    
9  &  $10\,471$  & 0.56\ ,\ 0.21
 	&  2.98 \\
 	
\midrule[1.8pt] 

Fit width & No.\ of quasar spectra &  \multicolumn{1}{c}{Redshift\ \ } &  \multicolumn{1}{c}{$\left[F_{\lambda}\times\sigma\right]_{5008
/4960}$} \\ \midrule[0.7pt] 
 
$2\sigma$  &  $10\,363$ & 0.56\ ,\ 0.21 
 	& 2.96 \\

$3\sigma$  &  $10\,252$  & 0.59\ ,\ 0.20
 	&  2.92 \\ 
 	
$4\sigma$  &  $9978$  & 0.59\ ,\ 0.20
 	& 2.98 \\ 
    
$5\sigma$  &  $9726$  & 0.56\ ,\ 0.21
 	&  2.98 \\
 	
\midrule[1.8pt] 

\end{tabular}

\end{table}

%There have been some attempts in the literature to detect a dependence of $\alpha$ on the gravitational potential using white dwarfs spectra \citep{Berengut}. In order to test any dependence on $\alpha$ on the gravitational potential in the narrow line region, we use the DR7 quasar catalogue from \citet{Yue} where the virial mass for the black hole has been estimated. There are many caveats on these virial masses \citep{MassBH}. We cross-match the quasars included in this catalogue with the DR12Q catalogue. There are 208 quasar spectra which are included in both catalogues. These spectra have a more complex continuum and display strong H\,$\beta$ lines. Thus, we apply the constraints described in Section \ref{sec:systematics} (vi) besides the mild constraints already discussed. We are left with 88 quasars, for which $\Delta\alpha/\alpha=\left(12\pm18\right)\times10^{-5}$. Further, we select two subsamples based on the black hole mass. The results are shown in Table \ref{tab:hemispheres}. At this stage we conclude that we do not have a statistical meaningful sample to test any dependence on the black hole mass.

\begin{table}

\caption{Results for SDSS-II/DR7 and BOSS (SDSS-III/DR12) samples. For each sample, the number of quasar spectra, the mean redshift, together with its standard deviation and value for $\Delta\alpha/\alpha$ are shown.}
\label{tab:dr7}
\centering
\begin{tabular}{@{}l@{\hspace{0.15cm}}c@{\hspace{0.15cm}}D{,}{\pm}{4.4}@{\hspace{0.15cm}}D{,}{\pm}{4.4}@{}}%{c | c | c } 
\midrule[1.8pt] 
Sample & No.\ of quasar spectra &  \multicolumn{1}{c}{Redshift\ \ \ \ } &  \multicolumn{1}{c}{$\Delta\alpha/\alpha\ (\times 10^{-5})$} \\ \midrule[0.7pt] 
 
DR7  &  $2853$ & 0.38\ ,\ 0.15
 	&  0.5\ ,\ 2.8  \\

DR7 (SDSS cont.)  &  $3009$  & 0.38\ ,\ 0.15
 	&  -0.4\ ,\ 2.7   \\ 
 	
BOSS (DR12)  &  $10\,363$  & 0.56\ ,\ 0.21
 	&  1.4\ ,\ 2.3  \\ 
    
BOSS + DR7  &  $13\,175$  & 0.51\ ,\ 0.21
 	&  0.9\ ,\ 1.8  \\
 	
\midrule[1.8pt] 

\end{tabular}\smallskip\\
%\subcaptionbox{$^{(a)}$ 10,363 (DR12 quasars) + 2,853 (DR7 quasars) - 41 (repeated quasars)}[\textwidth]
\end{table}

\begin{table}

\caption{Values of $\Delta\alpha/\alpha$ using the combined sample BOSS+DR7 for the same redshift bins as in Table \ref{tab:bins}.}
\label{tab:binscombined}
\centering
\begin{tabular}{@{}c@{\hspace{0.25cm}}c@{\hspace{0.2cm}}D{,}{\pm}{4.4}@{\hspace{0.15cm}}D{,}{\pm}{4.4}@{}}%{c | c | c } 
\midrule[1.8pt] 
Redshift interval & No.\ of quasar spectra &  \multicolumn{1}{c}{Redshift\ \ \ \ } &  \multicolumn{1}{c}{$\Delta\alpha/\alpha\ (\times 10^{-5})$} \\ \midrule[0.7pt] 
 
$0.390-0.460$ &  $1279$ & 0.42\ ,\ 0.02
 	&  -2.5\ ,\ 4.8  \\

$0.460-0.520$  &  $1076$  & 0.49\ ,\ 0.02
 	&  7.2\ ,\ 6.4  \\ 
 	
$0.520-0.580$  &  $1071$  & 0.55\ ,\ 0.02
 	&  1.1\ ,\ 7.1  \\ 
 	
$0.580-0.625$  &  $1025$  & 0.60\ ,\ 0.01
 	&  30.8\ ,\ 8.1  \\     
    
$0.625-0.675$  &  $1191$  & 0.65\ ,\ 0.01
 	&  -5.1\ ,\ 6.5 \\
 	
$0.675-0.715$  &  $1424$  & 0.69\ ,\ 0.01
 	&  -6.2\ ,\ 7.0 \\
 	
$0.715-0.765$  &  $1220$  & 0.74\ ,\ 0.01
 	&  1.4\ ,\ 6.8 \\
 	
$0.765-0.820$  &  $1519$  & 0.79\ ,\ 0.02
 	&  15.0\ ,\ 8.1 \\
 	
$0.820-0.880$  &  $644$  & 0.84\ ,\ 0.02
 	&  4.7\ ,\ 9.2  \\
 	
$0.880-1.000$  &  $580$  & 0.93\ ,\ 0.03
 	&  17.0\ ,\ 13.3  \\
 	
\midrule[1.8pt] 

\end{tabular}\smallskip\\
\end{table}

\begin{figure}
\centerline{\includegraphics[angle=0,width=0.49\textwidth]{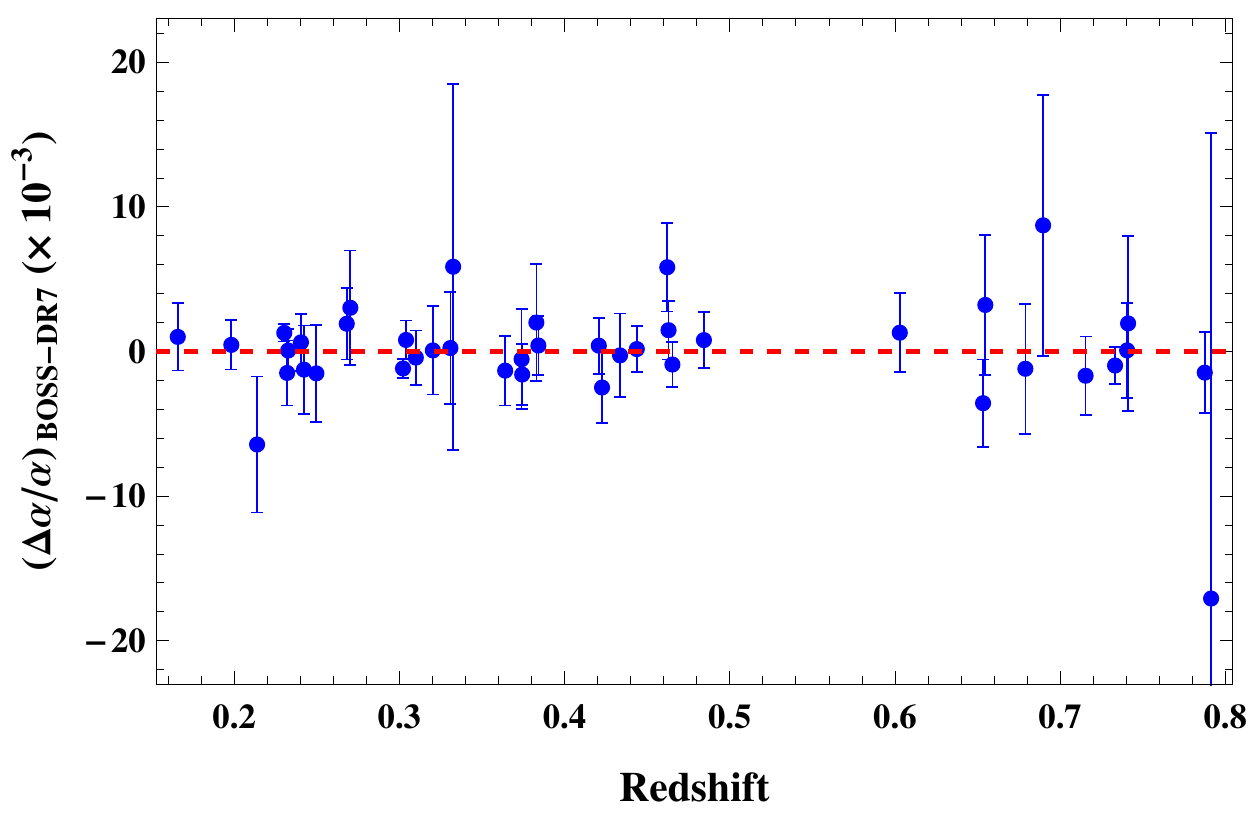}}
\caption{Difference between $(\Delta\alpha/\alpha)_{\rm{BOSS}}$ and $(\Delta\alpha/\alpha)_{\rm{DR}7}$ measurements for the 41 quasars observed by SDSS-I/II that were re-observed by BOSS. Both values for $\Delta\alpha/\alpha$ are consistent.
}
\label{fig:dr7}
\end{figure}

Finally, we have also considered quasar spectra from the SDSS-II/DR7, which were observed using the previous spectrograph instead of the upgraded BOSS spectrograph (see Section \ref{sec:sample}). From the DR7 quasar catalogue \citep{Schneider}, which contains 105\,783 quasars, we select a sample of 2853 quasars up to redshift $z=0.8$ using the same criteria described in Section \ref{sec:sample}. This number is similar to the quasar spectra considered by \citet[][Table \ref{tab:comparison}]{Rahmani}. We re-analyse this sample using the methodology presented in this work, and we find $\Delta\alpha/\alpha=\left(0.5\pm2.8\right)\times 10^{-5}$. By combining this DR7 sample with our fiducial BOSS (DR12) quasar sample (after eliminating 41 spectra which were re-observed by BOSS), we obtain our final sample which contains a total of 13\,175 quasars. The value obtained for this combined sample is reported as a final result of this investigation:
\begin{eqnarray}
\frac{\Delta\alpha}{\alpha}=\left(0.9 \pm 1.8 \right) \times 10^{-5} \nonumber\,.
\end{eqnarray}
Table \ref{tab:dr7} shows the results for DR7, DR7 using the continuum fit provided by the SDSS pipeline,\footnote{The SDSS pipeline provides a continuum fit for the DR7 spectra. The good agreement between the value for $\Delta\alpha/\alpha$ obtained with the SDSS continuum fit and our continuum fit is a good test for our code.} BOSS (DR12) and the combined BOSS+DR7. It can be seen that the mean redshift for the DR7 sample is lower than that for BOSS. Note that there is also a big difference on the mean S/N$_{\oxygen 5008}$ of both samples: S/N$_{\oxygen 5008}^{\rm{DR}7}=60$ and S/N$_{\oxygen 5008}^{\rm{BOSS}}=33$, which also explains why the statistical errors for $\Delta\alpha/\alpha$ do not reflect the expected reduction due to the increase in the number of quasars of our BOSS sample. Table \ref{tab:binscombined} shows the results of $\Delta\alpha/\alpha$ using the combined sample in the same redshift bins considered for our fiducial sample. Fig. \ref{fig:dr7} shows the difference of the values obtained for $\Delta\alpha/\alpha$ for the 41 re-observed quasars. Both BOSS and DR7 measurements are in perfect agreement within the error bars. This test is a good check for the reliability of our code and the consistency of the SDSS spectra obtained with different spectrographs.

\begin{figure}
\centerline{\includegraphics[angle=0,width=0.49\textwidth]{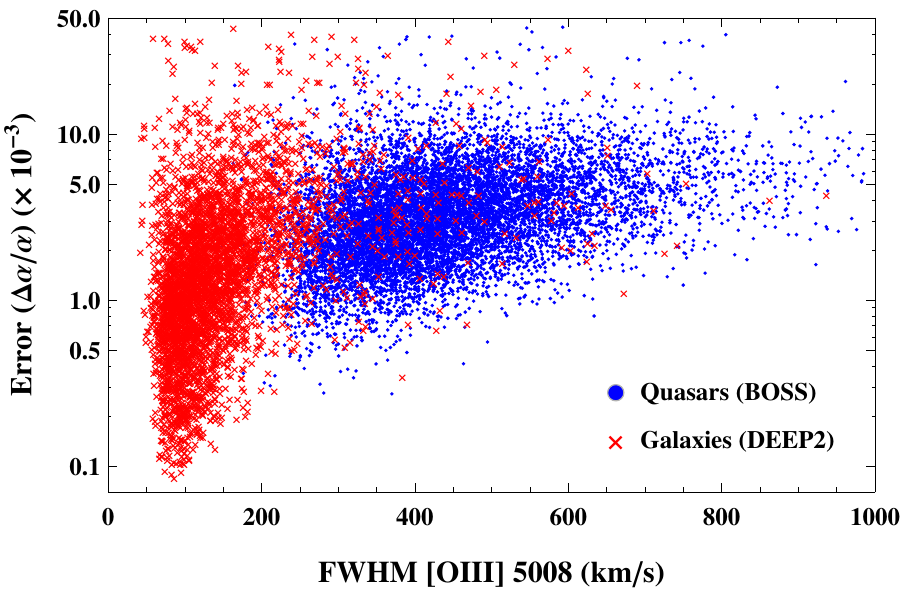}}
\caption{Errors on $\Delta\alpha/\alpha$ as a function of the FWHM of \oxygen 5008 for our fiducial BOSS quasar sample (blue points) and for the DEEP2 galaxy sample (red crosses).}
\label{fig:deep}
\end{figure}

There are massive galaxy surveys to be carried out during the next decade. For instance, eBOSS and DESI will take spectra from millions of galaxies. Therefore, it is interesting to give an estimation of the accuracy that will be obtained when using galaxy spectra instead of quasars. For this, we have analysed the galaxy spectra collected by the DEEP2 survey \citep{DEEP2} taken with resolving power $\approx6000$. From this sample, we found 4056 galaxies with strong [OIII] lines. Naively, one would expect that the error on $\Delta\alpha/\alpha$ should be
\begin{eqnarray}
 \Delta\left(\Delta\alpha/\alpha\right)_{\rm{galaxies, DEEP2}} \approx  f_{\rm{sample}}\times f_{\rm{inst}}\times f_{\rm{object}} \times \Delta\left(\Delta\alpha/\alpha\right)_{\rm{quasars, BOSS}}\,,
\end{eqnarray}
where $f_{\rm{sample}}(=\sqrt{10\,363/4056})$ takes into account the different number of objects in each sample, $f_{\rm{inst}}(\approx2000/6000)$ stands for the different resolution of the spectra and $f_{\rm{object}}$ is an extra factor due to the different characteristics of quasar and galaxy emission lines which affect the uncertainty of the line positions. This last factor is proportional to the line widths and inversely proportional to the line fluxes:
\begin{eqnarray}
f_{\rm{object}}\approx\frac{\rm{FWHM}_{\rm{galaxies}}}{\rm{FWHM}_{\rm{quasars}}}
\times\,\left(\frac{\rm{Flux}_{\rm{galaxies}}}{\rm{Flux}_{\rm{quasars}}}\right)^{-1}\,.
\end{eqnarray}
These numbers for the \oxygen 5008 line are approximately FWHM$_{\rm{galaxies}}\approx120$ km\,s$^{-1}$, FWHM$_{\rm{quasars}}\approx420$ km\,s$^{-1}$, Flux$_{\rm{galaxies}}\approx70$ and Flux$_{\rm{quasars}}\approx210$ in units of $10^{-17}$ erg\,cm$^{-2}$\,s$^{-1}$, obtained from the DEEP2 sample and from our fiducial sample. Thus, the expected error is $1.1\times10^{-5}$, where we have considered the error of our fiducial sample $\Delta\left(\Delta\alpha/\alpha\right)_{\rm{quasars, BOSS}}= 2.3\times10^{-5}$. Applying the same criteria described in Section \ref{sec:sample} to the DEEP2 galaxy sample, we get $\Delta\alpha/\alpha=\left(-0.9\pm1.6\right)\times 10^{-5}$. Thus, the upcoming future galaxy surveys will be quite competitive for constraining the variation of the fine-structure constant at low redshift $z<2$. Fig.\ \ref{fig:deep} shows the error on $\Delta\alpha/\alpha$ for the DEEP2 and BOSS samples.

\subsection{\neon lines}
\label{sec:neon}

\begin{figure}
\centerline{\includegraphics[angle=0,width=0.475\textwidth]{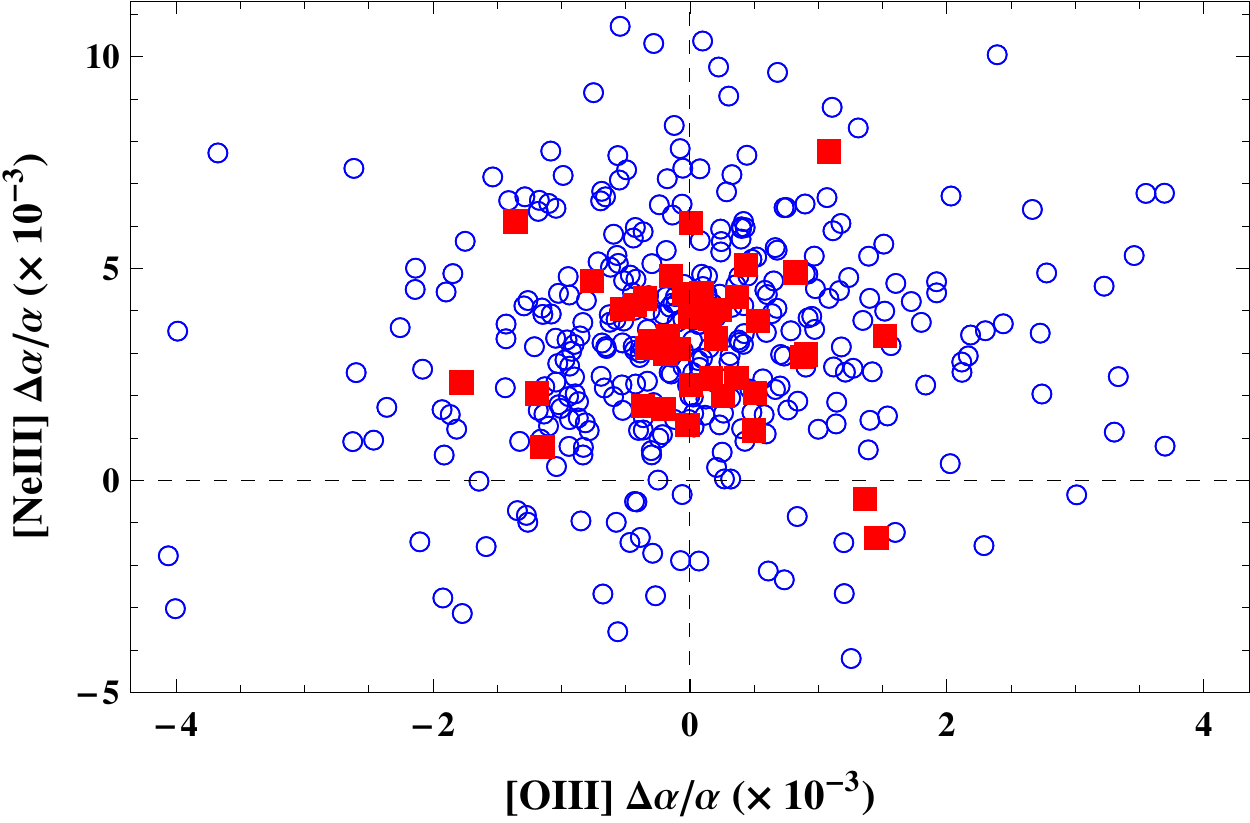}}
\caption{\neon and \oxygen measurements for $\Delta\alpha/\alpha$. Empty symbols stand for spectra with S/N$_{\rm{\neon 3869}}<35$ and solid squares for spectra with S/N$_{\rm{\neon 3869}}>35$. \neon measurements have a clear tendency to a positive variation of $\alpha$, which is due to a systematic effect affecting the \neon measurement. The same effect has already been noticed by \citet{Gutierrez}, and it is explained in Section \ref{sec:neon}.}
\label{fig:NeO}
\end{figure}

\begin{figure*}
\centerline{\includegraphics[angle=0,width=0.5\textwidth]{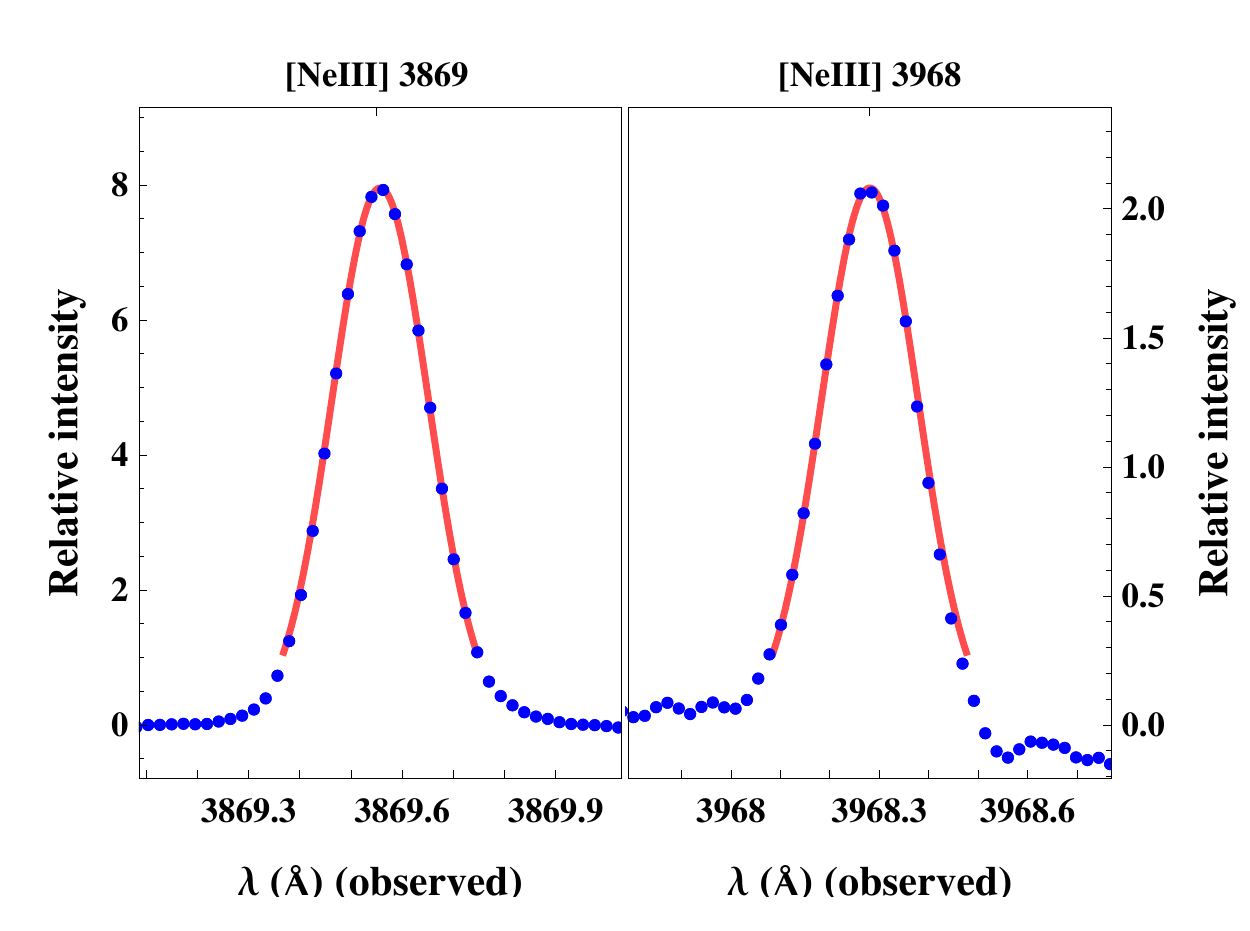}
\includegraphics[angle=0,width=0.492\textwidth]{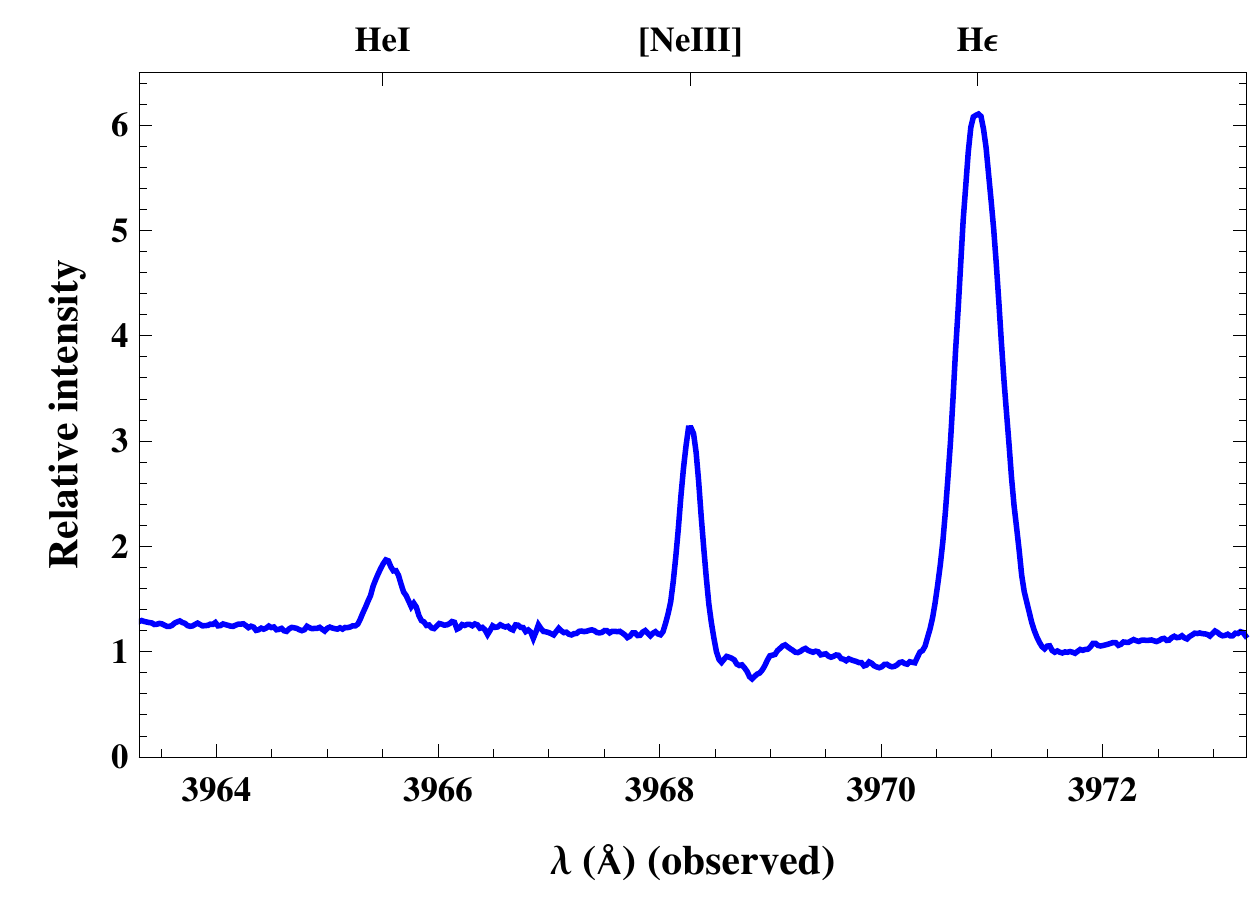}}
\caption{Left-hand panel: \neon 3869 and \neon 3968 together with our Gaussian fits (solid line) from a high-resolution ($R\approx25\,000$) spectrum of the Planetary Nebula IC 418 obtained with the FIES spectrograph at the NOT telescope. Right-hand panel: IC 418 spectrum centred at \neon 3968 line. The two close lines are  H\,$\epsilon$ 3971 \AA\ and He {\sc i} 3965 \AA.}
\label{fig:neon}
\end{figure*}

\begin{figure*}
\centerline{\includegraphics[angle=0,width=0.57\textwidth]{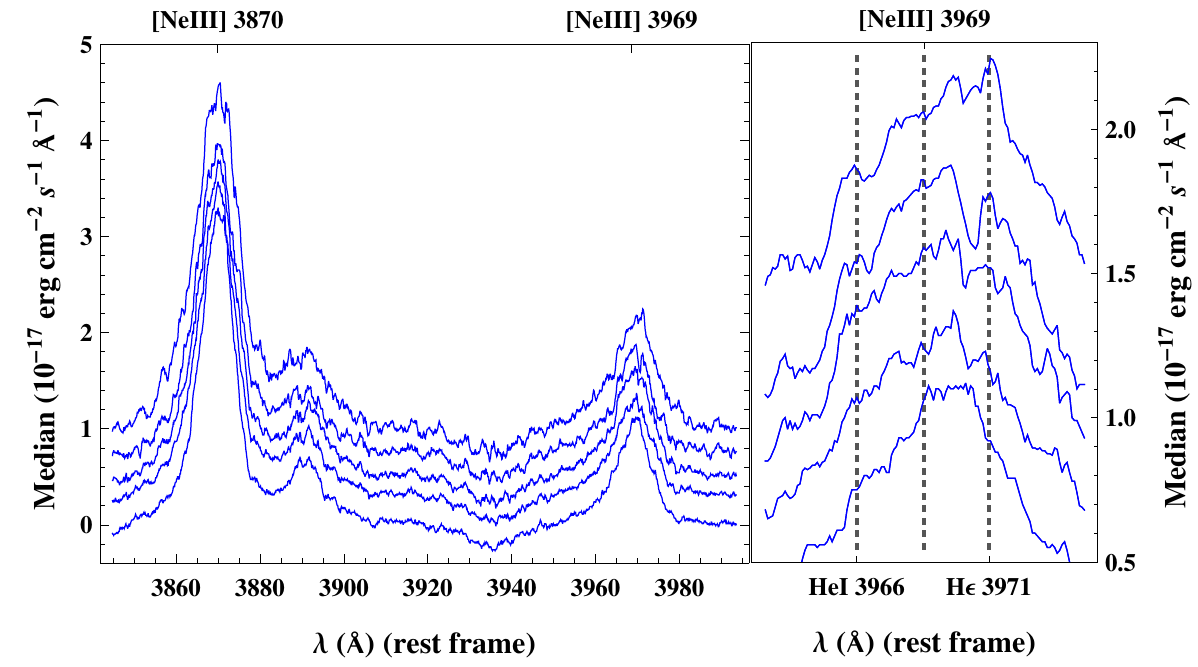}
\includegraphics[angle=0,width=0.405\textwidth]{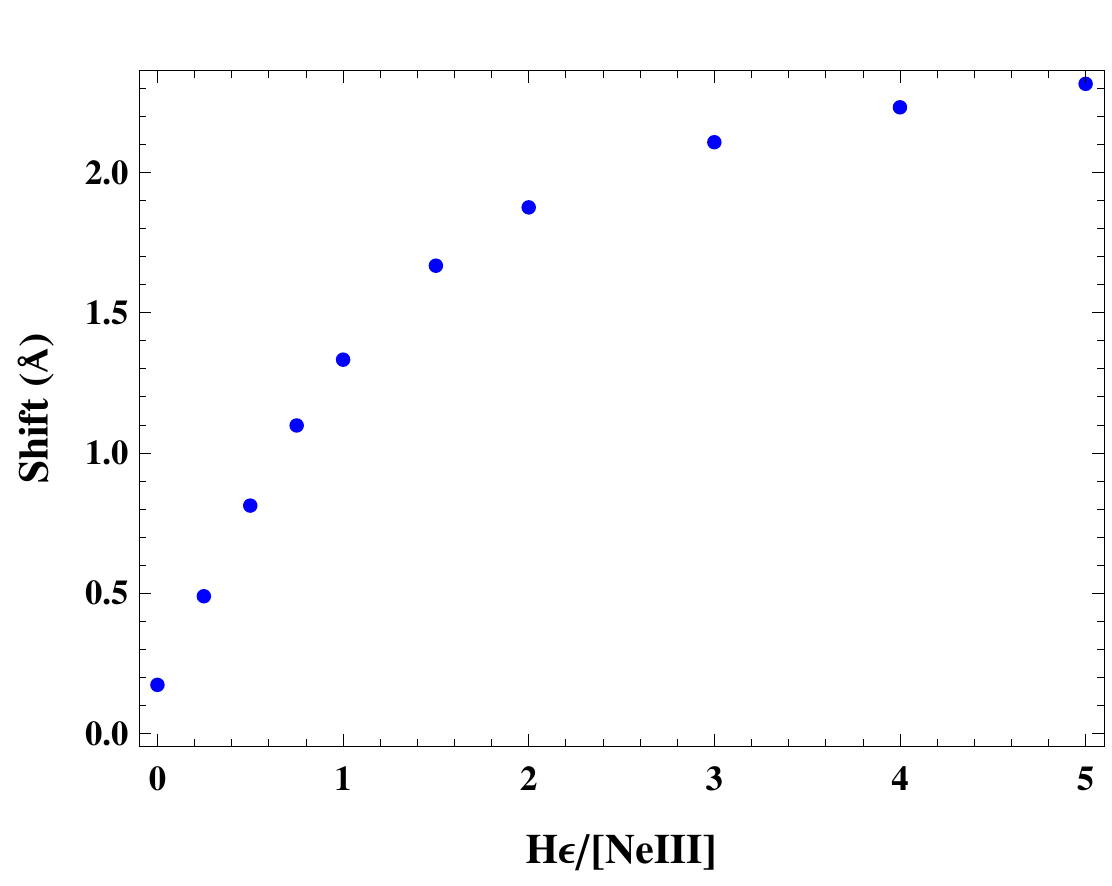}}
\caption{Left-hand panel: median-stacked quasar spectra with broad \neon lines (increasing line width from bottom to top). Both \neon lines are shown (\linesNe). The weak \neon line is blended with the two lines H\,$\epsilon$ 3971 \AA\ and He {\sc i} 3965 \AA. Right-hand panel: shift produced by the H\,$\epsilon$ line in the \neon 3968 line as a function of the line intensity ratio of both lines as measured from the Planetary Nebula convolved spectrum.}
\label{fig:blended}
\end{figure*}

We also measure from 462 quasar spectra with \neon emission lines the following constraint on the fine-structure constant:
\begin{eqnarray}
\Delta\alpha/\alpha_{\rm{\neon}}=\left(34\pm1\right)\times 10^{-4}\,,
\end{eqnarray} 
to be compared with
\begin{eqnarray}
\Delta\alpha/\alpha_{\rm{\neon}}=\left(36\pm1\right)\times 10^{-4}
\end{eqnarray}
obtained by \citet{Gutierrez}. The analysis of the \neon lines reveals the same systematic effect previously observed, namely a clear tendency for a positive variation of $\alpha$. \mbox{Fig.\ \ref{fig:NeO}} compares the results obtained for $\Delta\alpha/\alpha$ for spectra where both \oxygen and \neon lines are present. To account for this effect, a shift $\sim 0.6\,\AA$ on the theoretical or observed values of the  wavelengths for the \neon lines is necessary. There are experimental \citep{NeIIIlines} and indirect \citep{Kramida} values for the wavelengths of the \neon lines which are in agreement with errors $\approx3\times 10^{-2}\,\AA$. We use the NIST values for the \neon lines
\begin{eqnarray}
\lambda^{\rm{\neon}}_{1}=3869.86 \;\rm{\AA}\, \hspace{0.6cm} \lambda^{\rm{\neon}}_{2}=3968.59\;\rm{\AA}\,
\end{eqnarray}
\begin{eqnarray}
\delta\lambda^{\rm{\neon}}_{0}= 98.73 \;\rm{\AA}\,.
\end{eqnarray}
The results for the \oxygen doublet guarantee the good calibration of the SDSS spectra (and many more independent scientific results based on the SDSS spectra). Thus, we have measured the \neon lines using a high-resolution optical spectrum from
the planetary nebula IC 418. The IC 418 optical spectrum (∼3600-7200 \AA) was taken under service  
time at the Nordic Optical Telescope (NOT; Roque de los Muchachos, La Palma) in 2013 March with the FIES spectrograph. We used FIES in the low-resolution mode ($R\approx 25\,000$) with
the 2.5 arsec fibre (centred at the central star of IC 418). Three  
exposures of 1200\,s each
were combined into a final IC 418 spectrum, reaching a S/N (in the  
stellar continuum)
of $\sim$60 at 4000\ \AA\ and in excess of $\sim$150 at wavelengths longer
than 5000\ \AA\ \citep[see][for more observational details]{Anibal}. To measure $\Delta\alpha/\alpha$, we need to know the ratio
\begin{eqnarray}
{\cal R}\,=\,\left[\left(\lambda_{2}-\lambda_{1}\right)/\left(\lambda_{2}+
\lambda_{1}\right)\right]_{0}\,,
\end{eqnarray}
which is independent of the peculiar velocity of the planetary nebula. From our data, we obtain
\begin{eqnarray}
{\cal R}\,=\,\left(1259561\pm 4 \right)\times 10^{-8}\,,
\end{eqnarray}
compared to the one using NIST values for the wavelengths
\begin{eqnarray}
{\cal R}_{\rm{NIST}}\,=\, 1259560 \times 10^{-8}\,.
\end{eqnarray}
The difference between the two values translates into a variation on $\Delta\alpha/\alpha<10^{-6}$. Thus, the measured wavelength separation for the \neon doublet does not account for the positive variation on $\alpha$ observed using these lines. Fig.\ \ref{fig:neon} (left-hand panel) shows the Gaussian fit to the \neon line profiles present in the IC 418 spectrum.

The IC 418 spectrum shows two different lines near the \neon 3968 \AA\ line (see Fig. \ref{fig:neon}, right-hand panel). The stronger one is H\,$\epsilon$ 3971 \AA, the other one is \mbox{He {\sc i} 3965 \AA}. Hence, we search for a possible blending of the \neon line 3968 with these two lines in our much lower spectral resolution quasar spectra. Fig.\ \ref{fig:blended} (left-hand panel) shows stack quasar spectra with broad \neon emission lines. It can be seen that the weak \neon line is blended.

To quantify the displacement produced by the blending with H\,$\epsilon$ line, we did a Gaussian convolution of the Planetary Nebula spectrum to lower the resolution down to $R\approx2000$. Since the line intensity ratio of \neon and H\,$\epsilon$ may differ in the quasar narrow emission-line region and the Planetary Nebula, we show in Fig.\ \ref{fig:blended} (right-hand panel) the shift produced by the H\,$\epsilon$ line as a function of the ratio \neon/H\,$\epsilon$. We get a shift $\sim$0.6 \AA\ when H\,$\epsilon$/\neon is $\sim$0.5. This explains the systematic found when using \neon lines to measure the variation of the fine-structure constant $\Delta\alpha/\alpha$ in previous studies \citep{Gutierrez, Grupe}.

\vspace{-0.5cm}
%%%%%%%%%%%%%%%%%%%%%%%%%%%%%%%%%%%%%%%%%%%%%%%%%%%%%%
%%%%%%%%%%%%%%%%%%%%%%%%%%%%%%%%%%%%%%%%%%%%%%%%%%%%%%

\section{Summary}
\label{sec:summary}
%%%%%%%%%%%%%%%%%%%%%%%%%%%%%%%%%%%%%%%%%%%%%%%%%%%%%%
%%%%%%%%%%%%%%%%%%%%%%%%%%%%%%%%%%%%%%%%%%%%%%%%%%%%%%

The main conclusions of this work are as follows.

\begin{enumerate}

\item From 45\,802 objects at $z<1$ classified as quasars in the SDSS-III/BOSS DR12 quasar catalogue, we have extracted a sample of 10\,363 quasars with \oxygen emission lines. Combining this fiducial sample with a sample of 2853 previously studied SDSS-II/DR7 quasars, we got a final sample of 13\,175 after eliminating 41 re-observed quasars.\smallskip

\item With this combined sample, we have estimated a value for the possible variation of the fine-structure constant of \mbox{$\Delta\alpha/\alpha=\left(0.9 \pm 1.8\right)\times 10^{-5}$}, which represents the most accurate result obtained with this methodology.\smallskip

\item We have also studied how much our results change when analysing the fiducial sample according to different properties (width, amplitude, S/N and $R^2$ coefficient of the \oxygen lines), and when modifying some parameters of the analysis (polynomial order for the continuum subtraction, different methods to determine the line position, e.g. Gaussian/Voigt profiles). We conclude that our results are quite robust, and they are consistent with no variation of the fine-structure constant.\smallskip

%\item Systematic effects have been studied such as misidentification of the lines, H\,$\beta$ contamination, broad lines, width for the Gaussian fits, polynomial order for the continuum spectrum, S/N and the goodness of the fits. We have also compared the results with the ones obtained by using the different methods described in Section \ref{sec:methodology} to measure the position of the \oxygen lines and they are shown to be consistent.\smallskip

\item From over one million simulated realizations of quasar spectra, we conclude that the precision of our emission-line method is dominated by the error from the Gaussian fits. Hence, the error from the continuum subtraction and any possible systematics from our code are small.\smallskip

\item The standard deviation of the results as a function of redshift correlates with the sky. This result suggests that our main source of uncertainty is determined by the sky subtraction algorithm.\smallskip

\item We have determined the ratio of the \oxygen transition lines to be $2.96\pm0.02_{\rm syst}$, which is in good agreement with previous experimental and theoretical values.\smallskip

\item The same systematic effect previously noticed by \citet{Gutierrez} has been found on the \neon lines measurement. Incorrect measurement for the separation of the \neon has been excluded as a possible explanation, and a blending of the H\,$\epsilon$ and the \neon 3968 has been identified as the source of this effect.\smallskip

\item The measurement of $\Delta\alpha/\alpha$ using SDSS-III/BOSS spectra has reached the maximum precision unless better sky subtraction algorithms are developed. To obtain better constraints ($<10^{-6}$) using the emission-line method, high-resolution spectroscopy ($R\approx 100\,000$) is mandatory.\smallskip

\item We note that future large galaxies survey like eBOSS or DESI could provide quite stringent constraint for $\Delta\alpha/\alpha$ at low redshift, following our analysis of galaxy spectra taken from the DEEP2 survey.\smallskip

\end{enumerate}

\vspace{-0.5cm}
\section*{Acknowledgements}

FDA, JC, and FP acknowledge support from the Spanish MICINNs Consolider-Ingenio 2010 Programme under grant MultiDark CSD2009-00064, MINECO Centro de Excelencia Severo Ochoa Programme under grant SEV-2012-0249, and MINECO grants AYA-2012-31101 and AYA2014-60641-C2-1-P. FDA also acknowledges financial support from `la Caixa'-Severo Ochoa doctoral fellowship, UAM+CSIC Campus of International Excellence and Instituto de Astrof\'isica de Canarias for a summer stay where this work began. AM
and DAGH acknowledge support provided by the Spanish  Ministry of Economy and Competitiveness (MINECO) under grant AYA-2011-27754.

Funding for SDSS-III has been provided by the Alfred P.\ Sloan
Foundation, the Participating Institutions, the National Science
Foundation, and the U.S. Department of Energy Office of Science.
The SDSS-III web site is \href{http://www.sdss3.org/}{http://www.sdss3.org/}.

SDSS-III is managed by the Astrophysical Research Consortium for the Participating Institutions of the SDSS-III Collaboration including the University of Arizona, the Brazilian Participation Group, Brookhaven National Laboratory, University of
Cambridge, Carnegie Mellon University, University of Florida, the
French Participation Group, the German Participation Group, Harvard University, the Instituto de Astrof\'isica de Canarias, the Michigan State/Notre Dame/JINA Participation Group, Johns Hopkins
University, Lawrence Berkeley National Laboratory, Max Planck
Institute for Astrophysics, Max Planck Institute for Extraterrestrial Physics, New Mexico State University, New York University,
Ohio State University, Pennsylvania State University, University of
Portsmouth, Princeton University, the Spanish Participation Group,
University of Tokyo, University of Utah, Vanderbilt University,
University of Virginia, University of Washington, and Yale University.

This article is also partially based on service observations made with  
the Nordic Optical Telescope operated on the island of La Palma by the Nordic Optical Telescope Scientific Association in the Spanish Observatorio del Roque de Los Muchachos of the Instituto de Astrof\'isica de Canarias.

\appendix

\section*{Appendix A}
\renewcommand\thefigure{A\arabic{figure}} 
\setcounter{figure}{0} 
\renewcommand\thetable{A\arabic{table}} 
\setcounter{table}{0} 

We publish along with this paper an electronic table with the combined SDSS-III/BOSS DR12 and SDSS-II/DR7 sample of 13\,175 quasars used in this work. Table \ref{tab:table} describes the information and format of each column. The table is available in the following link \href{http://mnras.oxfordjournals.org/content/suppl/2015/08/11/stv1406.DC1/suppl_data.zip}
{http://mnras.oxfordjournals.org/content/suppl/2015/
08/11/stv1406.DC1/suppl\_data.zip}.

\newcolumntype{L}[1]{>{\raggedright\let\newline\\\arraybackslash\hspace{0pt}}m{#1}}
\newcolumntype{C}[1]{>{\centering\let\newline\\\arraybackslash\hspace{0pt}}m{#1}}
\newcolumntype{R}[1]{>{\raggedleft\let\newline\\\arraybackslash\hspace{0pt}}m{#1}}
\begin{table*}

\caption{Description of the electronic table with the combined sample (13\,175 quasars) published along with the paper.}
\label{tab:table}
\centering
\begin{tabularx}{1.0\textwidth}{L{1cm}L{2cm}C{2cm}L{10cm}}%{c | c | c } 
\midrule[1.8pt] 
Column & Name &  Format &  Description \\ \midrule[0.7pt] 
1 & \verb|SDSS_NAME| & STRING & SDSS-DR12 designation hhmmss.ss+ddmmss.s (J2000) \\
2 & \verb|RA| & DOUBLE & Right Ascension in decimal degrees (J2000) \\
3 & \verb|DEC| & DOUBLE & Declination in decimal degrees (J2000) \\
4 & \verb|THING_ID| & INT32 & Thing\verb|_|ID\\
5 & \verb|PLATE| & INT32 & Spectroscopic plate number\\
6 & \verb|MJD| & INT32 & Spectroscopic MJD ($>$55\,000 SDSS-III/BOSS spectra, $<$55\,000 SDSS-II spectra)\\
7 & \verb|FIBER| & INT32 & Spectroscopic fiber number\\
8 & \verb|Z_VI| & DOUBLE & Redshift from visual inspection\\
9 & \verb|Z_PIPE| & DOUBLE & Redshift from BOSS pipeline\\
10 & \verb|ERR_ZPIPE| & DOUBLE & Error on BOSS pipeline redshift\\
11 & \verb|ALPHA| & FLOAT & $\Delta\alpha/\alpha$ from the Gaussian fits\\
12 & \verb|ERR_ALPHA| & FLOAT & Standard error for $\Delta\alpha/\alpha$ from the Gaussian fits\\
13 & \verb|SN_O1| & FLOAT & S/N for the \oxygen 4960 line\\
14 & \verb|SN_O2| & FLOAT & S/N for the \oxygen 5008 line\\
15 & \verb|O1_FIT| & FLOAT & Line centroid for the \oxygen 4960 line\\
16 & \verb|O2_FIT| & FLOAT &  Line centroid for the \oxygen 5008 line\\
17 & \verb|ERR_O1| & FLOAT & Error on the line centroid for the \oxygen 4960 line\\
18 & \verb|ERR_O2| & FLOAT & Error on the line centroid for the \oxygen 5008 line\\
19 & \verb|O1_AMPLITUDE| & FLOAT & Gaussian amplitude at the centre for the \oxygen 4960 line\\
20 & \verb|O2_AMPLITUDE| & FLOAT & Gaussian amplitude at the centre for the \oxygen 5008 line\\
21 & \verb|O1_WIDTH| & FLOAT & Gaussian width for the \oxygen 4960 line\\
22 & \verb|O2_WIDTH| & FLOAT & Gaussian width for the \oxygen 5008 line\\
23 & \verb|FILE_NAME| & STRING & File name to download from the SDSS server\\

\midrule[1.8pt] 

\end{tabularx}

\end{table*}

\vspace{-0.5cm}
\bibliographystyle{mn2e}
\bibliography{biblio.bib}

\label{lastpage}

\end{document}